\documentclass[longauth]{aa}

\usepackage{subfigure}
\usepackage{float}
\usepackage{txfonts}
\usepackage{graphicx}
\usepackage{natbib}
\usepackage{xspace}
\usepackage{color}

\bibpunct{(}{)}{;}{a}{}{,}

\def\({\left(}
\def\){\right)}
\def\[{\left[}
\def\]{\right]}
\def\mhmpc{\,h^{-1}{\rm Mpc}}

\def\mhgpcc{\,h^{-3}{\rm Gpc^3}}

\newcommand*\pdt{P_{\delta\theta}\(k\)}
\newcommand*\ptt{P_{\theta\theta}\(k\)}

\newcommand{\kms}{$\,{\rm km\, s^{-1}}$\xspace}

\def\bigstrutup{\vrule width0pt height0.3truecm depth0truecm}
\def\bigstrutdown{\vrule width0pt height0truecm depth0.16truecm}

\DeclareMathOperator{\popcnt}{popcnt}

\begin{document}

\title{The VIMOS Public Extragalactic Redshift Survey (VIPERS)\thanks{
    Based on observations collected at the European Southern
    Observatory, Cerro Paranal, Chile, using the Very Large Telescope
    under programs 182.A-0886 and partly 070.A-9007.  Also based on
    observations obtained with MegaPrime/MegaCam, a joint project of
    CFHT and CEA/DAPNIA, at the Canada-France-Hawaii Telescope (CFHT),
    which is operated by the National Research Council (NRC) of
    Canada, the Institut National des Sciences de l’Univers of the
    Centre National de la Recherche Scientifique (CNRS) of France, and
    the University of Hawaii. This work is based in part on data
    products produced at TERAPIX and the Canadian Astronomy Data
    Centre as part of the Canada-France-Hawaii Telescope Legacy
    Survey, a collaborative project of NRC and CNRS. The VIPERS web
    site is http://www.vipers.inaf.it/.}  }

\subtitle{Unbiased clustering estimate with VIPERS slit assignment}

\titlerunning{Redshift-space distortions in VIPERS}

\author{
F.~G.~Mohammad\inst{\ref{brera},\ref{unimi}}
\and D.~Bianchi\inst{\ref{icg}}
\and W.~J.~Percival\inst{\ref{uw},\ref{pi},\ref{icg}}
\and S.~de la Torre\inst{\ref{lam}}
\and L.~Guzzo\inst{\ref{unimi},\ref{brera}}
\and B.~R.~Granett\inst{\ref{brera},\ref{unimi}}
\and E.~Branchini\inst{\ref{roma3},\ref{infn-roma3},\ref{oa-roma}}
%
%
\and M.~Bolzonella\inst{\ref{oabo}}      
\and B.~Garilli\inst{\ref{iasf-mi}}          
\and M.~Scodeggio\inst{\ref{iasf-mi}}       
%
%
\and U.~Abbas\inst{\ref{oa-to}}
\and C.~Adami\inst{\ref{lam}}
\and J.~Bel\inst{\ref{cpt},\ref{brera}}
\and D.~Bottini\inst{\ref{iasf-mi}}
\and A.~Cappi\inst{\ref{oabo},\ref{nice}}
\and O.~Cucciati\inst{\ref{unibo},\ref{oabo}}         
\and I.~Davidzon\inst{\ref{lam},\ref{oabo}}   
\and P.~Franzetti\inst{\ref{iasf-mi}}   
\and A.~Fritz\inst{\ref{iasf-mi}}       
\and A.~Iovino\inst{\ref{brera}}
\and J.~Krywult\inst{\ref{kielce}}
\and V.~Le Brun\inst{\ref{lam}}
\and O.~Le F\`evre\inst{\ref{lam}}
\and K.~Ma{\l}ek\inst{\ref{warsaw-nucl}},\inst{\ref{lam}}
\and F.~Marulli\inst{\ref{unibo},\ref{infn-bo},\ref{oabo}} 
\and M.~Polletta\inst{\ref{iasf-mi},\ref{marseille-uni},\ref{toulouse}}
\and A.~Pollo\inst{\ref{warsaw-nucl},\ref{krakow}}
\and L.A.M.~Tasca\inst{\ref{lam}}
\and R.~Tojeiro\inst{\ref{st-andrews}}  
\and D.~Vergani\inst{\ref{iasf-bo}}
\and A.~Zanichelli\inst{\ref{ira-bo}}
%
%
\and S.~Arnouts\inst{\ref{lam},\ref{cfht}} 
\and J.~Coupon\inst{\ref{geneva}}
\and G.~De Lucia\inst{\ref{oats}}
\and O.~Ilbert\inst{\ref{lam}}
\and L.~Moscardini\inst{\ref{unibo},\ref{infn-bo},\ref{oabo}}
\and T.~Moutard\inst{\ref{halifax}}
}
\institute{
INAF - Osservatorio Astronomico di Brera, Via Brera 28, 20122 Milano --  via E. Bianchi 46, 23807 Merate, Italy \label{brera}
\and  Universit\`{a} degli Studi di Milano, via G. Celoria 16, 20133 Milano, Italy \label{unimi}
\and Institute for Cosmology and Gravitation, University of Portsmouth, 1-8 Burnaby Rd, Portsmouth PO1 3, UK \label{icg}
\and  Department of Physics and Astronomy, University of Waterloo, 200 University Ave W, Waterloo, ON N2L 3G1, Canada \label{uw}
\and Perimeter Institute for Theoretical Physics, 31 Caroline St. North, Waterloo, ON N2L 2Y5, Canada \label{pi}
\and Aix Marseille Univ, Universit\'e Toulon, CNRS, CPT, Marseille, France \label{cpt}
\and Dipartimento di Matematica e Fisica, Universit\`{a} degli Studi Roma Tre, via della Vasca Navale 84, 00146 Roma, Italy\label{roma3} 
\and INFN, Sezione di Roma Tre, via della Vasca Navale 84, I-00146 Roma, Italy \label{infn-roma3}
\and INAF - Osservatorio Astronomico di Roma, via Frascati 33, I-00040 Monte Porzio Catone (RM), Italy \label{oa-roma}
\and Aix Marseille Univ, CNRS, CNES, LAM, Marseille, France  \label{lam}
\and Dipartimento di Fisica e Astronomia - Alma Mater Studiorum Universit\`{a} di Bologna, via Gobetti 93/2, I-40129 Bologna, Italy \label{unibo}
\and INFN, Sezione di Bologna, viale Berti Pichat 6/2, I-40127 Bologna, Italy \label{infn-bo}
\and INAF - Osservatorio Astronomico di Bologna, via Gobetti 93/3, I-40129, Bologna, Italy \label{oabo} 
\and Institute for Astronomy, University of Edinburgh, Royal Observatory, Blackford Hill, Edinburgh EH9 3HJ, UK \label{roe}
\and INAF - Istituto di Astrofisica Spaziale e Fisica Cosmica Milano, via Bassini 15, 20133 Milano, Italy \label{iasf-mi}
\and INAF - Osservatorio Astrofisico di Torino, 10025 Pino Torinese, Italy \label{oa-to}
\and Laboratoire Lagrange, UMR7293, Universit\'e de Nice Sophia Antipolis, CNRS, Observatoire de la C\^ote d’Azur, 06300 Nice, France \label{nice}
\and Institute of Physics, Jan Kochanowski University, ul. Swietokrzyska 15, 25-406 Kielce, Poland \label{kielce}
\and National Centre for Nuclear Research, ul. Hoza 69, 00-681 Warszawa, Poland \label{warsaw-nucl}
\and Aix-Marseille Université, Jardin du Pharo, 58 bd Charles Livon, F-13284 Marseille cedex 7, France \label{marseille-uni}
\and IRAP,  9 av. du colonel Roche, BP 44346, F-31028 Toulouse cedex 4, France \label{toulouse} 
\and Astronomical Observatory of the Jagiellonian University, Orla 171, 30-001 Cracow, Poland \label{krakow} 
\and School of Physics and Astronomy, University of St Andrews, St Andrews KY16 9SS, UK \label{st-andrews}
\and INAF - Istituto di Astrofisica Spaziale e Fisica Cosmica Bologna, via Gobetti 101, I-40129 Bologna, Italy \label{iasf-bo}
\and INAF - Istituto di Radioastronomia, via Gobetti 101, I-40129,Bologna, Italy \label{ira-bo}
\and Canada-France-Hawaii Telescope, 65--1238 Mamalahoa Highway, Kamuela, HI 96743, USA \label{cfht}
\and Department of Astronomy, University of Geneva, ch. d’Ecogia 16, 1290 Versoix, Switzerland \label{geneva}
\and INAF - Osservatorio Astronomico di Trieste, via G. B. Tiepolo 11, 34143 Trieste, Italy \label{oats}
\and Department of Astronomy \& Physics, Saint Mary's University, 923 Robie Street, Halifax, Nova Scotia, B3H 3C3, Canada \label{halifax}
}

\authorrunning{Mohammad et al.}

\offprints{\mbox{F.~G.~Mohammad},\\ \email{faizan.mohammad@brera.inaf.it}}

\abstract{
The VIPERS galaxy survey has measured the clustering of $0.5<z<1.2$ galaxies, enabling a number of measurements of galaxy properties and cosmological redshift-space distortions (RSD). Because the measurements were made using one-pass of the VIMOS instrument on the Very Large Telescope (VLT), the galaxies observed only represent approximately 47\% of the parent target sample, with a distribution imprinted with the pattern of the VIMOS slitmask. Correcting for the effect on clustering has previously been achieved using an approximate approach developed using mock catalogues. Pairwise inverse probability (PIP) weighting has recently been proposed to correct for missing galaxies, and we apply it to mock VIPERS catalogues to show that it accurately corrects the clustering for the VIMOS effects, matching the clustering measured from the observed sample to that of the parent. We then apply PIP-weighting to the VIPERS data, and fit the resulting monopole and quadrupole moments of the galaxy two-point correlation function with respect to the line-of-sight, making measurements of RSD. The results are close to previous measurements, showing that the previous approximate methods used by the VIPERS team are sufficient given the errors obtained on the RSD parameter.   
}

\maketitle

\section{Introduction} \label{sec:intro}                

The clustering of galaxies within galaxy surveys provides a wealth of astrophysical information, allowing measurements of galaxy formation, galaxy evolution, and cosmological parameters. Missing galaxies within surveys can however distort the clustering compared to that of the full population of the type of objects to be observed if the missed galaxies are not randomly chosen but instead cluster in a different way to the full population. Such a situation is often induced by the mechanics of the experimental apparatus, which, given a parent population of targets, limits what can actually be observed. In this paper we consider missing galaxies in the VIPERS survey \citep{Guzzo14,scodeggio16}. VIPERS collected $89022$ galaxy redshifts over an overall area of 23.5\,deg$^2$, covering the W1 and W4 fields of the Canada-France-Hawaii Telescope Legacy Survey Wide (CFHTLS-Wide){\footnote{{\tt http://www.cfht.hawaii.edu/Science/CFHTLS/}}}. A colour pre-selection was used to remove galaxies at $z<0.5$, helping to bring the sampling efficiency to 47\%. VIPERS conducted observations using the VIMOS multi-object spectrograph \citep{lefevre03}, which applies a slit-mask to select targets for follow-up spectroscopy. A brief description of VIPERS is provided in Section~\ref{sec:survey}.

The requirement that spectra taken with VIMOS should not overlap on the focal plane limits the placement of slits, and consequently the galaxies that can be observed. This effect is stronger along the dispersion direction compared to across it, because of the rectangular nature of the projected spectra. The occulted region around each galaxy is imprinted on the statistical distribution of the observed galaxies. There are no overlapping observations, such as those present in the Baryon Oscillation Spectroscopic Survey (BOSS \citealt{dawson16}), meaning that the lost information cannot be recovered: we simply do not have clustering information on scales smaller than the minimum separation perpendicular to the dispersion direction. On larger scales, the slitmask still impacts on the measured clustering through the large-scale pattern imprinted on the sky, and the density dependence of the selection.

 \citet{bianchi17} and \citet{percival17} presented a new method to correct for missing galaxies in surveys. This builds up a probability for each pair of galaxies in the observed sample to have been observed in a set of realisations of the survey\footnote{With the term `survey realisation' we indicate a possible outcome of the spectroscopic observation given an underlying parent sample. It is not to be confused with the term `survey mocks' that are built from an ensemble of parent catalogues keeping the observational setup fixed.}. These realisations, drawn from the same underlying parent catalogue, are all equally likely. Each sample can be obtained by simply re-running the targeting algorithm after moving or rotating the parent sample, or changing any random selection performed by the selection algorithm. We observe one of these sets of galaxies, and by inverse weighting by the pairwise probability of observation we force the clustering of the one realisations to match that of the set as a whole. Provided there are no pairs of zero weight, this weighting leads to a clustering estimate of the observed sample that is unbiased compared to that of the full parent sample. The method is described in more detail in Section~\ref{sec:pip}.

In this paper we apply this method to remove the effects of the VIMOS slitmask from the VIPERS survey. The slitmask has a strong effect, leading to an observed clustering signal that is very different from that expected \citep{delatorre13a}. In previous VIPERS papers this was approximately corrected using a target sampling rate (TSR) given by the fraction of potential targets placed behind a slit in a rectangular region around each targeted galaxy \citep{pezzotta16}. A further correction that up-weights galaxy pairs by the ratio $[1+w_s(\theta)]/[1+w_p(\theta)]$ of the angular clustering of the observed $w_s$ and parent $w_p$ samples \citep{delatorre13a} was also used to improve the small-scale clustering measurements. While similar in principle to the method of \citet{bianchi17} and \citet{percival17}, this relies on the missed pairs being statistically identical to the population as a whole. This is not the case in VIPERS as galaxies are more likely to be missed in denser regions where they have different properties. The TSR up-weighting method was extensively tested in past VIPERS analyses to provide a sub-percentage-level accuracy on the clustering measurements in mock catalogues. However, the TSR weighting is a parametric method that was calibrated on mock catalogues to minimise the systematic bias of the clustering measurements. It does not take into account possible differences in the clustering of simulated and observed datasets. The pairwise inverse probability (PIP) weighting scheme uses the data themselves to infer the selection probabilities providing the same level of accuracy. In this sense the new correction method is self-contained and more robust than the method based on using the TSR.

To optimise the design of the slitmasks, VIPERS uses the so-called {\tt SPOC} algorithm (Slit Positioning and Optimisation Code), within the ESO VIMOS mask preparation software {\tt VMMPS} \citep{bottini05}. {\tt SPOC} was designed to obtain the most spectra possible given an input parent sample. Rather than trying to change the internal properties of SPOC to make our set of realisations of the survey, we instead rely on spatially moving the survey mask and rotating the sample. We still miss all pairs that have a separation that is less than the minimum slit separation scale, but this is not an issue as we only consider larger scales here.

We use mock catalogues of VIPERS to test the new algorithm in Section~\ref{sec:obs_syst}, showing that it works as expected. Having corrected for the slit-mask effects, we consider how this changes the redshift-space distortions (RSD) signal within the sample. VIPERS was designed with RSD as one of the key measurements to be made: RSD are caused by the peculiar velocities of galaxies, which systematically distort redshifts leaving an enhanced clustering signal along the line-of-sight \citep{kaiser87}. By measuring the clustering anisotropy around the line-of-sight through observations of the multipole moments of the correlation function one can constrain the growth rate of cosmological structure parameterised by $f\sigma_8$, which constitutes the first-order contribution to the RSD signal. 

Early RSD measurements from VIPERS were based on the Public Data Release 1 sample \citep{garilli14}, measuring $f\sigma_8(z=0.8)$ \citep{delatorre13a}. Subsequent measurements from the final data sample, Public Data Release 2 \citep[PDR2][]{scodeggio16}, were presented by \citep{pezzotta16}. Extensions to these measurements include a configuration space joint analysis of RSD and weak-lensing \citep{delatorre16}, and an analysis splitting the sample based on galaxy type in order to extract extra information by comparing samples that trace the dark matter field in different ways \citep{mohammad17}. 

We present RSD measurements made by the `standard' two-point correlation function-based method in Section~\ref{sec:rsd_data}. These are compared to the previous VIPERS measurements, and we show that previous slit-mask-correction techniques were sufficient to make these measurements from VIPERS. This is discussed further in Section~\ref{sec:conc}.

To analyse the VIPERS-PDR2 data we used the same fiducial cosmology adopted in previous VIPERS clustering analyses, that is, a flat $\Lambda$CDM cosmology with parameters\\$(\Omega_b, \Omega_m, h, ns, \sigma_8)=(0.045, 0.3, 0.7, 0.96, 0.80)$.

\section{The VIPERS survey}     \label{sec:survey}                

\begin{figure}
        \centering
                \includegraphics[scale=0.5]{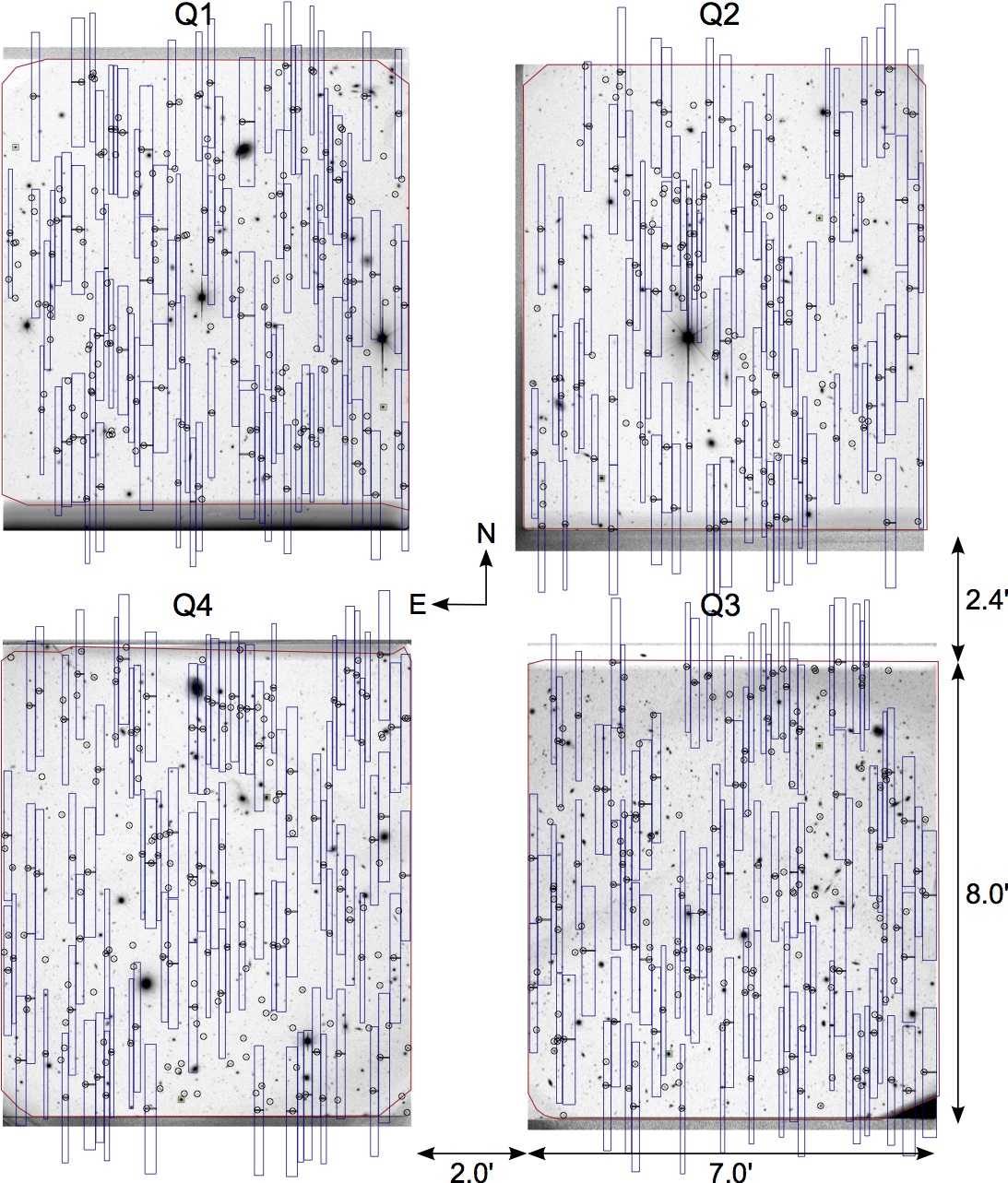}
                \caption{
        Example of the slit/spectrum distribution over a full VIMOS pointing, showing the disposition of the four quadrants and the "cross" among them. The circles identify the targets selected by the {\tt SPOC} optimisation algorithm. The elongated blue rectangles reproduce the "shadow" of the 2D spectrum that will result from each target in the final spectroscopic exposure. The thin red lines show the boundary of the actual spectroscopic mask, traced pointing-by-pointing through an automatic
detection algorithm that follows the borders of the illuminated area \citep[see][for details]{Guzzo14}. 
        }
        \label{fig:VIMOS-quadrant}
        \end{figure}

The VIPERS survey extends over an area of $ 23.5$ deg$^2$ within the W1 and W4 fields of the Canada-France-Hawaii Telescope Legacy Survey Wide (CFHTLS-Wide).  The VIMOS multi-object spectrograph \citep{lefevre03} was used to cover these two fields with a mosaic of 288 pointings, 192 in W1 and 96 in W4. Given the VIMOS footprint, which consists of four distinct quadrants separated by an empty "cross" of about 2 arcmin width (see Figure~\ref{fig:VIMOS-quadrant}), the survey area includes a regular grid of gaps where no galaxies were observed (see following section). Target galaxies were selected from the CFHTLS-Wide catalogue to a faint limit of $i_{\rm AB}=22.5$, applying an additional $(r-i)$ versus $(u-g)$ colour preselection that efficiently and robustly removes galaxies at $z<0.5$. Coupled with a highly optimised observing strategy \citep{scodeggio09}, this doubles the mean galaxy sampling efficiency in the redshift range of interest compared to a purely magnitude-limited sample, bringing it to 47\%.

Spectra were collected at moderate resolution ($R\simeq 220$) using the LR Red grism, providing a wavelength coverage of 5500-9500$\smash{\mathrm{\AA}}$. The typical redshift error for the sample of reliable redshifts is $\sigma_z=0.00054(1+z)$, which corresponds to an error on a galaxy peculiar velocity at any redshift of 163~\kms. These and other details are given in the PDR-2 release paper \citep{scodeggio16}. A discussion of the data reduction and management infrastructure was presented in \citet{garilli14}, while a complete description of the survey design and target selection was given in \citet{Guzzo14}.  The dataset used here  is the same early version of the PDR-2 catalogue used in \citet{pezzotta16} and \citet{delatorre16}, from which it differs by a few hundred redshifts revised during the very last period before the release. In total it includes $89\,022$ objects with measured redshifts. As in all statistical analyses of the VIPERS data, only measurements with quality flags 2 to 9 (inclusive) are used, corresponding to a sample with a redshift confirmation rate of $96.1\%$ \citep[for a description of the quality flag scheme, see][]{scodeggio16}. 


The procedures for defining the target list within the VIMOS spectroscopic masks were described in detail in \citet{bottini05}.  Within the {\tt VMMPS} environment, the {\tt SPOC} algorithm is used to optimise the position, size and, total number of slits. The final solution is derived by cross-correlating the user target catalogue with the corresponding object positions in a VIMOS direct exposure of the field (“pre-image”), observed beforehand. This operation matches the astrometric coordinates to the actual instrument coordinate system, selecting the subset of objects that will eventually deliver a spectrum and, potentially, a redshift measurement.  

{\tt SPOC} aims at finding an optimal disposition of the slits, packing the largest possible number of spectra over each quadrant (See \citealt{bottini05} for a detailed description of {\tt SPOC}). This happens irrespectively of the parent sample angular clustering. As such, it will tend to build a distribution that is more homogeneous on the sky compared to the full galaxy population at the corresponding magnitude limit. The denser the parent galaxy sample, the stronger the bias. If the number density of galaxies on the sky is much larger than the maximum density of slits that can be packed, {\tt SPOC} will essentially pick galaxies in a regular grid, packing the spectra in regular rows on top of each other.  This is not quite the case for VIPERS, for which the relatively bright magnitude limit allows for targeting, on average, about one half of the available galaxies, as shown in Figure \ref{fig:VIMOS-quadrant}. In this way, the measured sample still preserves a significant fraction of the original angular clustering.  Still, a bias is inevitably introduced and needs to be properly accounted for in any clustering measurement, which is the subject of this paper.  In addition, the finite size of slits introduces a proximity effect that also needs to be corrected for when computing galaxy clustering.

\begin{figure*}
        \centering
                \includegraphics[scale=0.28]{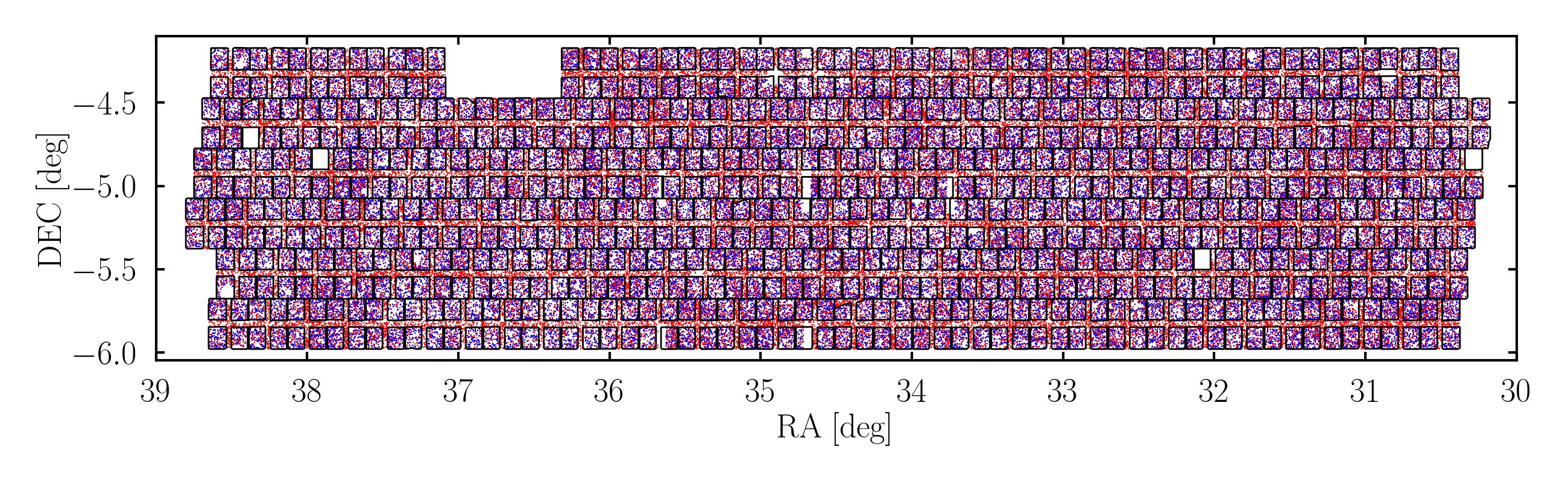}
        \includegraphics[scale=0.28]{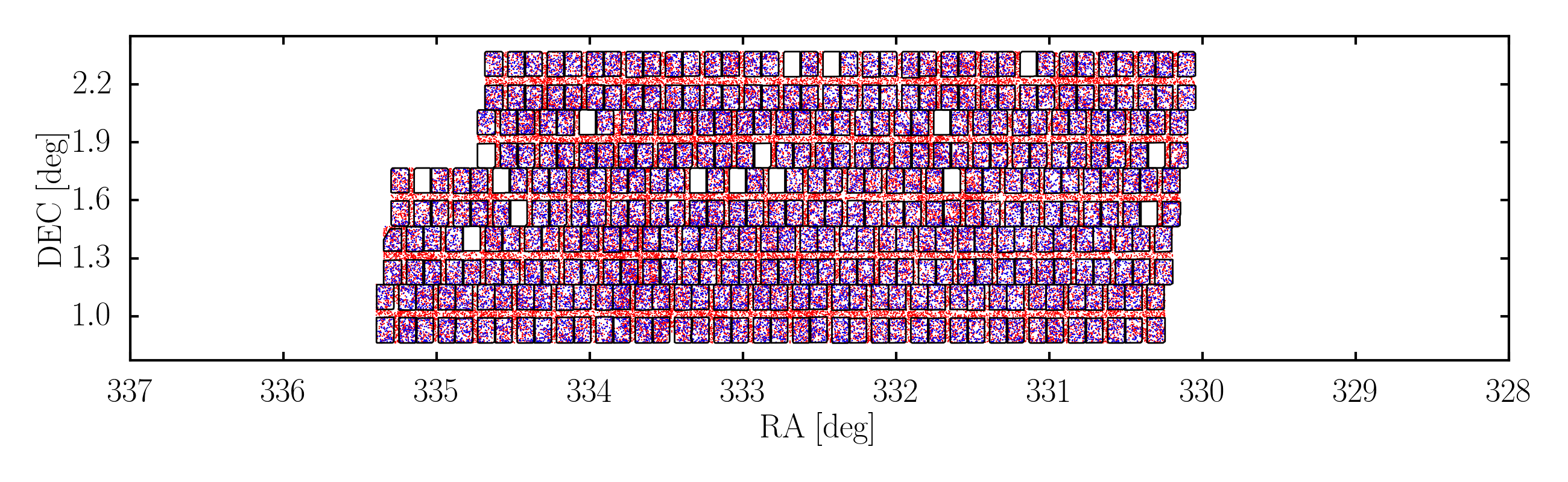}
                \caption{Scatter plot in the $(\text{RA},\text{DEC})$ plane for galaxies in the parent sample (red dots) and VIPERS-PDR2 catalogue (blue dots). Top and bottom panels show the W1 and W4 fields, respectively. Portions of the sky unobserved in the spectroscopic samples due to defects in the photometric sample, bright stars, or missing quadrants have been ascribed to the photometric mask.}
        \label{fig:ra-dec_scatter}
        \end{figure*}

Figure \ref{fig:VIMOS-quadrant} shows an example VIMOS observation. The overall mosaic of such pointings composing the full VIPERS survey is shown in Figure \ref{fig:ra-dec_scatter} for the two survey areas, W1 and W4. The boundaries of each single observation are described by the black polygons. In this figure, galaxies in the photometric parent sample and in the final VIPERS-PDR2 redshift catalogue are over-plotted as red  and blue dots, respectively. The gaps of the VIMOS footprint are clearly visible as vertical and horizontal stripes, in which only unobserved objects, marked in red, are present. In addition, the overall survey mask includes: (a) gaps in the photometric sample due to bright star or photometric problems (small irregular empty regions); (b) fully failed quadrants due to mechanical failure in the VIMOS metal mask insertion before the spectroscopic observation (white regular rectangles, mostly in W4); and (c) specific details in the spectroscopic observations, such as, for example, vignetting by the VLT guide probe \citep[described by the red line in Figure~\ref{fig:VIMOS-quadrant} -- see][for details]{Guzzo14,scodeggio16}

Throughout this work we have defined, as{ parent catalogue}, the photometric catalogue selected according to the VIPERS target selection function \citep{Guzzo14}, including all galaxies matching the external boundaries of the VIPERS-PDR2 sample, but with no mask applied. We have also ascribed the empty pointings and quadrants in the VIPERS-PDR2 sample to the photometric mask to avoid unnecessary complications in the implementation of the pipeline used for this analysis.

\section{VIPERS Mocks}  \label{sec:mocks}                                                

VIPERS mocks are based on the Big MultiDark Planck \citep[BigMDPL,][]{klypin16} dark matter N-body simulation. The simulation was carried out in the flat $\Lambda$CDM cosmological model with parameters: 
$\(\Omega_m,\Omega_b,h,n_s,\sigma_8 \)=
\(0.307,0.048,0.678,0.96,0.823\)$. 
Since the resolution is not sufficient to match the typical halo masses probed by VIPERS, low-mass haloes were added following the recipe proposed by \citet{delatorre13b}. 

Dark-matter halos were populated with galaxies using halo occupation distribution prescriptions with parameters calibrated using luminosity-dependent clustering measurements from early VIPERS  data. We refer the reader to \citet{delatorre13a,delatorre16} for a detailed description of the procedure. 

We used a set of 153 independent realistic parent and VIPERS-like mocks for each of the two VIPERS fields, W1 and W4. VIPERS-like mocks were obtained from the corresponding set of parent mocks in two steps: first, VIPERS targeting algorithm was applied by means of SPOC using the grid of VIPERS pointings; afterwards the footprint of VIPERS spectroscopic and photometric masks was imprinted to include the effect of obscured sky regions and quadrant vignetting (see Sect. \ref{sec:survey}). We also included the effect of VIPERS redshift error in the mock catalogues by blurring the cosmological redshifts using a Gaussian distribution of width $\sigma_z/(1+z)=0.00047$. Although different from the latest estimate from the PDR2 data, we used this value to perform a fair comparison of our results with those in \citet{pezzotta16}.

We used this set of mock samples to test the reliability of the weighting schemes proposed in \citet{bianchi17} and \citet{percival17}. The same set of mocks was also employed to estimate the data covariance matrix and quantify the systematic bias on estimates of the growth rate of structure.

\section{Measurements}  \label{sec:meas}                                           

\begin{table}
        \scriptsize
                \begin{center}
                \begin{tabular}{        c                                       c                                                       c                                                                       c}
                \hline
                \hline
                                                        Redshift        &               $N_{\mathrm{gal}}$              &               $V\ [\mhgpcc]$              &       $z_{\mathrm{eff}}$      \bigstrutup\bigstrutdown\\
                                \hline
                                $0.5<z<0.7$     &                       $30,910$                        &               $1.76\times10^{-3}$     &       $0.60$                          \bigstrutup\\
                                                        $0.7<z<1.2$     &                       $33,679$                        &               $7.34\times10^{-3}$     &       $0.86$                          \\
                                \hline
                                \hline
                \\
                        \end{tabular}
                \caption{Parameters characterising the two VIPERS subsamples split by redshift as used in this work. `Redshift' denotes the redshift range, $N_{gal}$ is the number of galaxies, $V$ stands for the volume in the VIPERS fiducial cosmology (see Sec. \ref{sec:intro}) and $z_{\mathrm{eff}}$ is the effective redshift of the sample computed as the median of the mean redshifts of galaxy-galaxy pairs with separations $5\mhmpc<s<50\mhmpc$. All figures refer to the full VIPERS, that is, both W1 and W4 fields.}\label{tab:z-split_data}
                \end{center}
        \end{table}

We measured the anisotropic two-point correlation function $\xi(s,\mu)$ as a function of the angle-averaged pair separation $s$ and the cosine $\mu$ of the angle between the pair separation and the line of sight. We employed the minimum variance Landy-Szalay estimator \citep{landy93},
        \begin{equation}
                \xi\(s,\mu\) = \frac{\mathrm{DD}\(s,\mu\)-2~\mathrm{DR}\(s,\mu\)}{\mathrm{RR}\(s,\mu\)}+1\; ,                                       \label{eq:LS}
        \end{equation}
where $\mathrm{DD}$, $\mathrm{DR,}$ and $\mathrm{RR}$ are the data-data, data-random, and random-random normalized pair counts, respectively. We binned $\mu$ in 200 linear bins in the range $0\le\mu\le1$ taking the mid-point of each bin as reference. The pair separation $s$ is instead binned using logarithmic bins,
        \begin{equation}
                \log s_{i+1} = \log s_i + \Delta s_{\log}\; ,                   \label{eq:log_bin}
        \end{equation}
with $\Delta s_{\log}=0.1$. The measurement in each pair separation bin is referenced to the logarithmic mean,
        \begin{equation}
                \log\langle s_i\rangle = \frac{\log s_i + \log s_{i+1}}{2}\; .                                                                                                                                       \label{eq:log_bin1}
        \end{equation}

The multipole moments $\xi^{s,\(\ell\)}\(s\)$ of the two-point correlation function are defined as its projection on the Legendre polynomials $L_{\ell}\(\mu\)$. Since we deal with discrete bins of the variable $\mu$, we replaced the integral by the Riemann sum such that,
        \begin{equation}
                \xi^{s,\(\ell\)}\(s_i\) = \left(2\ell+1\right)\sum_{j=1}^{200}\xi^s(s_i,\mu_j)L_{\ell}(\mu_j)\Delta\mu\ .                                               \label{eq:meas_mps}
        \end{equation}
When performing the angular pair counts $DD_a(\theta)$ and $DR_a(\theta)$ we used 100 linear bins within $0^\circ\le\theta\le8^\circ$. This range is sufficiently large to cover a transverse pair separation of\\$\sim185\mhmpc$ at $z=0.5$ in VIPERS fiducial cosmology.

Following \cite{pezzotta16} we divided the redshift range $0.5< z<1.2$ covered by VIPERS into two bins spanning $0.5< z<0.7$ and $0.7< z<1.2$ with effective redshifts of $z_{\mathrm{eff}}=0.60$ and $z_{\mathrm{eff}}=0.86,$ respectively. The subsample at low redshifts contains 30,910 galaxies while the one at high redshifts includes 33,679 galaxies. These parameters are listed in Table~\ref{tab:z-split_data}. Since VIPERS targeting over W1 and W4 fields was performed using the same observational setup we treated them as a single survey and performed the pair counts simultaneously on both fields rather than combining the measurements of the correlation function from each field.

\subsection{Mitigating for missing targets} \label{sec:pip}

The PIP approach provides us with unbiased estimates of the galaxy pair counts in the presence of missing observations, with the only formal requirement being that no pair has zero probability of being observed \citep{bianchi17}.

At each separation $\vec{s}$, the data-data pair counts are obtained as
\begin{equation}\label{eq PIPauw}
  DD(\vec{s}) = \sum_{\vec{x}_m - \vec{x}_n \approx \vec{s}} \mathrm{w}_{mn} \frac{DD^{(p)}_a(\theta)}{DD_a(\theta)} \ ,
\end{equation}
where $\mathrm{w}_{mn}=1/p_{mn}$ is the inverse of the selection probability of the
pair formed by the galaxies $m$ and $n$, whereas $DD^{(p)}_a$ and $DD_a$ represent the angular pair counts of parent and observed sample, respectively.
The observed angular pair counts are, in turn, computed via the same $\mathrm{w}_{mn}$ weights,
\begin{equation}\label{eq PIPauw ang}
  DD_a(\theta) = \sum_{\vec{u}_m \cdot \vec{u}_n \approx \cos(\theta)} \mathrm{w}_{mn} \ .
\end{equation}
For brevity, we have adopted the notation $\sum_{\vec{x}_m - \vec{x}_n \approx \vec{s}}$ and
$\sum_{\vec{u}_m \cdot \vec{u}_n \approx \cos(\theta)}$, with
$\vec{u}_i = \vec{x}_i/|\vec{x}_i|$, to indicate that the sum is
performed in bins of $\vec{s}$ and $\theta$, respectively.
Similarly, for the data-random pair counts,
\begin{equation}\label{eq PIPauw DR}
DR(\vec{s}) = \sum_{\vec{x}_m - \vec{y}_n \approx \vec{s}} \mathrm{w}_m \frac{DR^{(p)}_a(\theta)}{DR_a(\theta)} \ ,
\end{equation}
where $\mathrm{w}_m=1/p_m$ is the inverse of the selection probability of the galaxy
$m$, and
\begin{equation}\label{eq PIPauw DR ang}
  DR_a(\theta) = \sum_{\vec{u}_m \cdot \vec{v}_n \approx \cos(\theta)} \mathrm{w}_m \ .
\end{equation}

We evaluate the selection probabilities $p_{mn}$ and $p_m$ empirically, by creating an ensemble of possible outcomes of the target selection given an underlying parent catalogue; that is, we rerun  the slit-assignment algorithm on the same parent sample several times (see Sect. \ref{sec:pipeline}).
As discussed in \citet{bianchi17}, rather than storing all the PIP weights (one for each pair), it is convenient to compress the information in the form of individual bitwise weights (one for each galaxy).
The bitwise weight of a galaxy $\mathrm{w}_i^{(b)}$ is defined as a binary array, of length $N_{runs}$, in which the $n$-th bit equals 1 if the galaxy has been selected in the $n$-th targeting realisation and 0 otherwise.
$N_{runs}$ represents, by construction, the total number of realisations. 
For convenience, we use base-ten integers to encode the bitwise weights.
The PIP weights are obtained `on the fly', while doing pair counts, as
\begin{equation}\label{eq wb DD}
  \mathrm{w}_{mn} = \frac{N_{runs}}{\popcnt\left[\mathrm{w}^{(b)}_m \ \& \ \mathrm{w}^{(b)}_n \right]} \ ,
\end{equation}
where $\&$ and $\popcnt$ are fast bitwise operators, which multiply two integers bit by bit and return
the sum of the bits of the resulting integer, respectively.
Similarly, for individual weights, we have
\begin{equation}\label{eq wb DR}
  \mathrm{w}_m = \frac{N_{runs}}{\popcnt\left[\mathrm{w}^{(b)}_m \right]} \ .
\end{equation}

The requirement that all pairs are observable (they can be observed in at least one VIPERS realisation) means that the expectation value of the PIP estimator (excluding angular up-weighting) matches the clustering of all of the pairs within the parent sample - those targeted for possible VIPERS observation. Pairs in the parent sample that cannot be observed would formally have infinite weight but, practically, they would never appear in the pair counts in a particular realisation of VIPERS\footnote{For the sake of clarity, we note that $S_{\text{pairs}} \subseteq S_{\text{observable pairs}} \subseteq S_{\text{observed pairs}}$, where $S_x$ stands for set of $x$. We also note that, in general, it is not possible to infer $S_{\text{observable pairs}}$ from $S_{\text{observed galaxies}}$.}.
If we have some pairs that are not observable (they have zero probability of observation), angular up-weighting can serve two different purposes:\\
(i) The number of unobservable pairs is not negligible, but the clustering of these pairs is statistically equivalent to that of the observable ones. This happens when being observable or not is a property that does not depend on clustering; for example, when galaxies fall in a blind spot of the instrument's focal plane (see \citealt{Bianchi18} for a more detailed discussion). In this case it is formally correct to use the full set of observable plus unobservable pairs to perform angular upweighting to recover unbiased estimates of the three-dimensional clustering. We note that here these regions would not be excluded in the mask used to create the random catalogue.\\
(ii) The unobservable pairs are such because of their clustering but the total number is small enough that their effect is negligible, at least on the scales of interest. In this second scenario angular upweighting is simply a way to reduce the variance and the more self-consistent approach is to use only the set of observable pairs. Using the full set of pairs could potentially increase the effect of the unobservable pairs. \\
As discussed in Sect. \ref{sec:obs_syst}, the VIPERS survey is compatible with category (ii).
Interestingly, we find that the mean fraction of unobservable pairs in mock samples is about a factor of two larger than what is shown in Fig. \ref{fig:un-targ_frac} for the VIPERS-PDR2 galaxy sample. This points to some difference between mocks and data in terms of galaxy clustering.
Unlike the weighting schemes calibrated on simulated datasets (e.g. TSR weights), PIP weights are built to be insensitive to this difference.
We use mocks just to verify that the effect of unobservable pairs is confined to the smallest scales.
The above mentioned factor two guaranties that the same conclusion holds for real data.
\\ 

\subsection{Correcting for redshift failures}\label{sec:ssr}

The reliability of each VIPERS redshift measurement is quantified by a quality flag. Spectroscopic redshift measurements with a quality flag 2 to 9 (inclusive) have a redshift confirmation rate of $96.1\%$ and are regraded as reliable. We label all objects that do not satisfy this condition as `redshift failures'. The reliability of a redshift measurement depends on a number of factors such as the field-to-field observational conditions and the presence of clear spectral features and presents a correlation with some galaxy properties such as colour and luminosity. The effect of redshift failures is quantified by means of the spectroscopic success rate (SSR) defined as the ratio between the number of objects with a reliable redshift measurement (in our case the ones with a quality flag between 2 and 9) and the total number of targets placed behind a slit in a given VIMOS quadrant. It is computed as a function of the galaxy rest-frame U-V colour and B-band luminosity and is assigned to each galaxy with a reliable redshift measurement.

To correct the clustering measurements against redshift failures, we have up-weighted each galaxy by the corresponding weight $\mathrm{w}_{\mathrm{SSR}}=\mathrm{SSR}^{-1}$. Equations \eqref{eq PIPauw} and \eqref{eq PIPauw ang} are therefore modified as
        \begin{equation}\label{eq:ssrdd}
                DD(\vec{s}) = \sum_{\vec{x}_m - \vec{x}_n \approx \vec{s}} \mathrm{w}_{mn} \frac{DD^{(p)}_a(\theta)}{DD_a(\theta)} \times \mathrm{w}_{\mathrm{SSR}}^{(m)}\mathrm{w}_{\mathrm{SSR}}^{(n)} \; ,
        \end{equation}
and
        \begin{equation}\label{eq:ssrdd_ang}
                DD_a(\theta) = \sum_{\vec{u}_m \cdot \vec{u}_n \approx \cos(\theta)} \mathrm{w}_{mn} \times \mathrm{w}_{\mathrm{SSR}}^{(m)}\mathrm{w}_{\mathrm{SSR}}^{(n)}\; .
        \end{equation}
Data-random cross-pair counts in Eqs. \eqref{eq PIPauw DR} and \eqref{eq PIPauw DR ang} now become,
        \begin{equation}\label{eq:ssrdr}
DR(\vec{s}) = \sum_{\vec{x}_m - \vec{y}_n \approx \vec{s}} \mathrm{w}_m \frac{DR^{(p)}_a(\theta)}{DR_a(\theta)} \times \mathrm{w}_{\mathrm{SSR}}^{(m)}\ ,
\end{equation}
and
        \begin{equation}\label{eq:ssrdr_ang}
                DR_a(\theta)=\sum_{\vec{u}_m\cdot\vec{v}_n \approx \cos(\theta)} \mathrm{w}_m \times \mathrm{w}_{\mathrm{SSR}}^{(m)}\ ,
        \end{equation}
respectively. 

The effect of redshift failures is not reproduced in the mock catalogues. We therefore make use of spectroscopic success rates only when dealing with the VIPERS-PDR2 galaxy catalogue.

\section{Pipeline} \label{sec:pipeline}
 \begin{figure}
        \centering
                \includegraphics[scale=0.20]{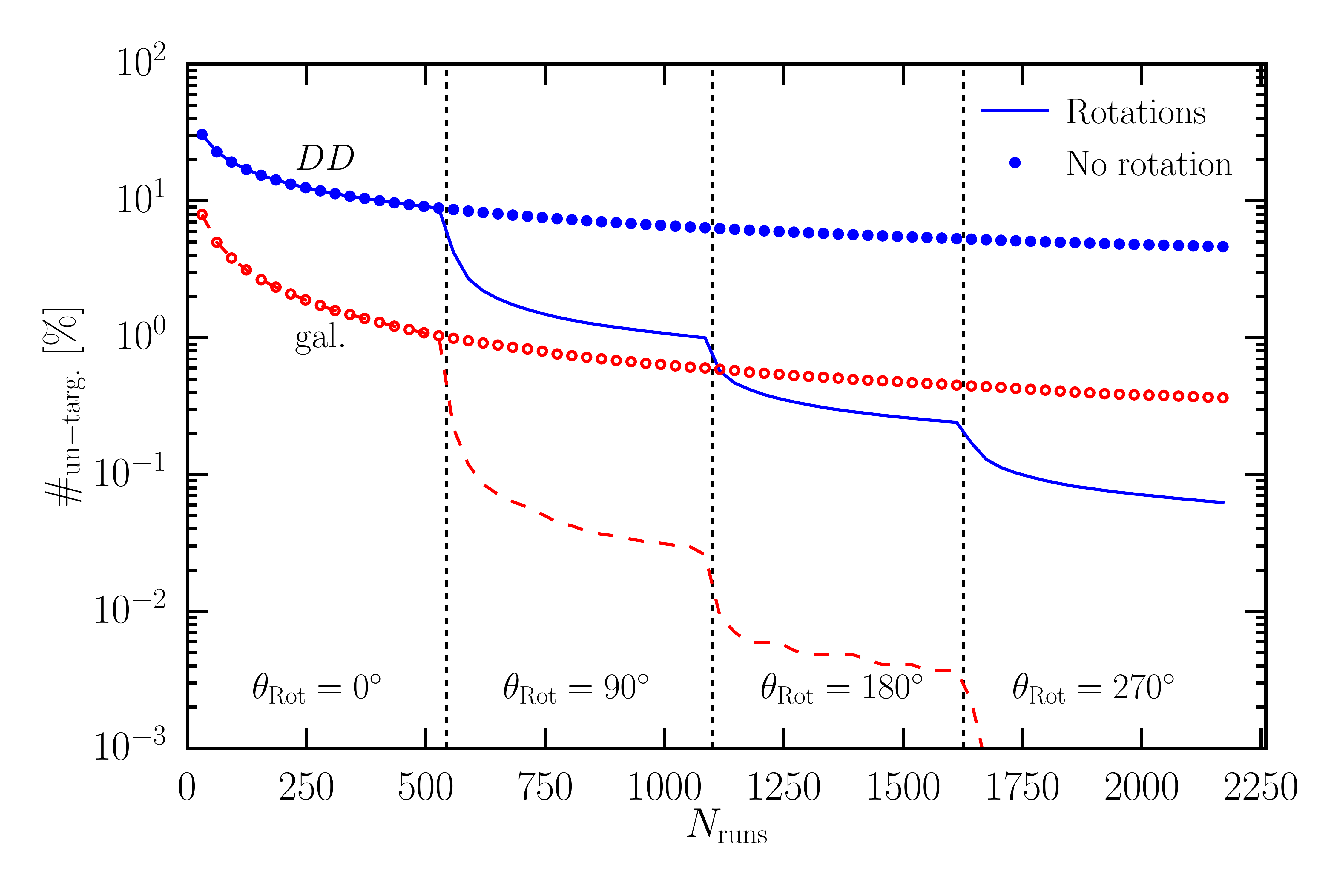}
                \caption{Fraction of unobservable galaxies and pairs of galaxies in the VIPERS parent catalogue as a function of the number of targeting runs $N_{\mathrm{runs}}$. Points show the case when multiple survey realisations are generated only spatially moving the spectroscopic mask, while lines result from also rotating the parent catalogue by $\theta_\mathrm{Rot}$. For the latter case, vertical dashed lines delimit the subset of targeting runs sharing the same $\theta_\mathrm{Rot}$. Blue filled points and continuous line show the fraction of unobservable galaxy-galaxy pairs while red empty markers and dashed line correspond to individual galaxies.}\label{fig:un-targ_frac}
        \end{figure}

 \begin{figure*}
        \centering
                \includegraphics[scale=0.15]{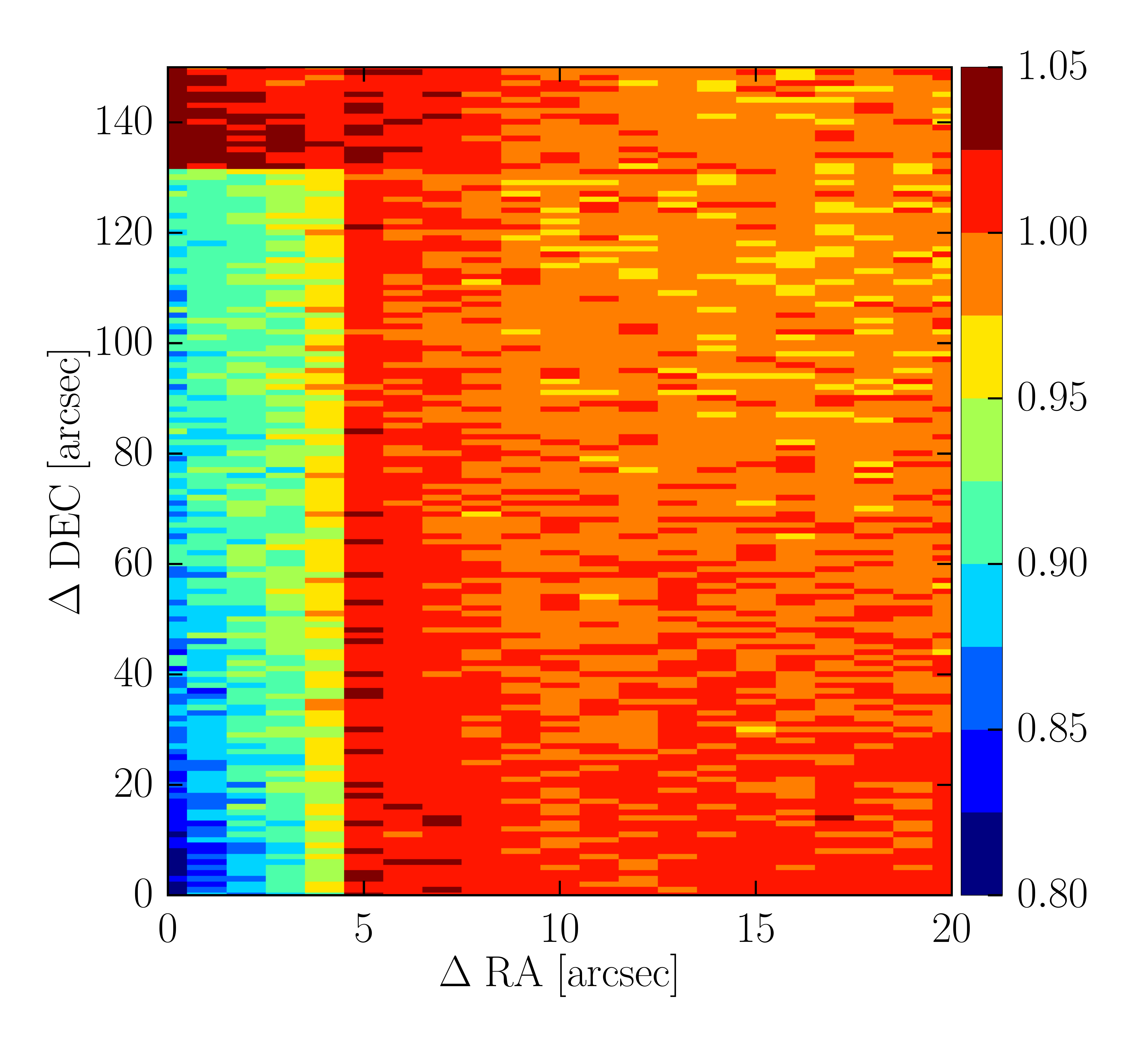}
        \includegraphics[scale=0.15]{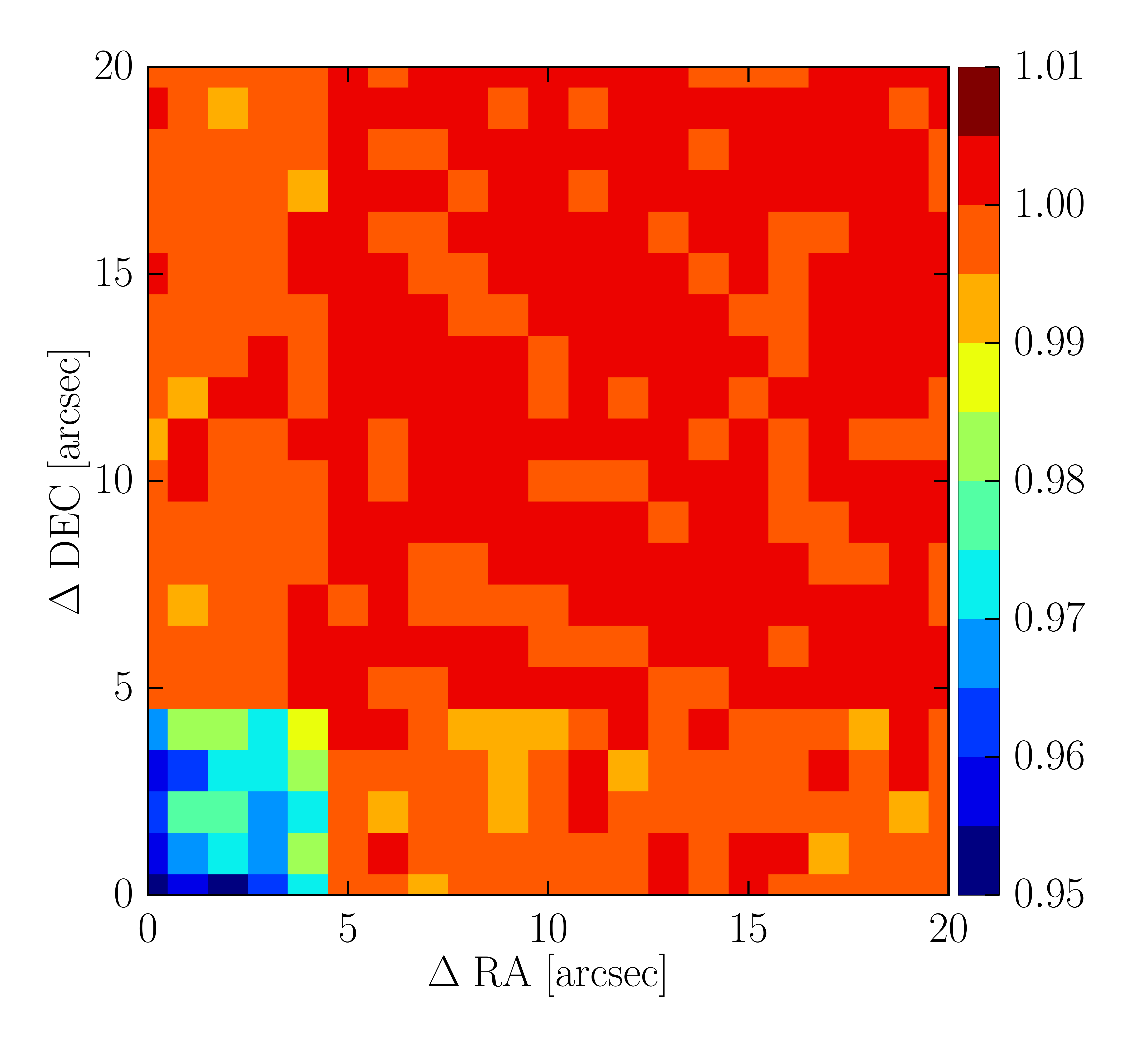}
                \caption{2D angular completeness function of galaxy-galaxy pairs (see Eq. \eqref{eq:completeness}) observed in 2170 survey realisations with respect to the VIPERS parent catalogue. Left panel: survey realisations are obtained spatially moving the spectroscopic mask over the parent catalogue. The rectangular `shadow' at small separations is the typical footprint of VIMOS spectra. Right panel: As in the left panel but when also the rotations of the parent catalogue are added to make multiple survey realisations. The size of the square shadow at small pair separations in the right panel is the typical length of VIMOS slits and is produced by the slit collisions only.}\label{fig:ang-compl}
        \end{figure*}

\begin{figure*}
        \centering
                \includegraphics[scale=0.26]{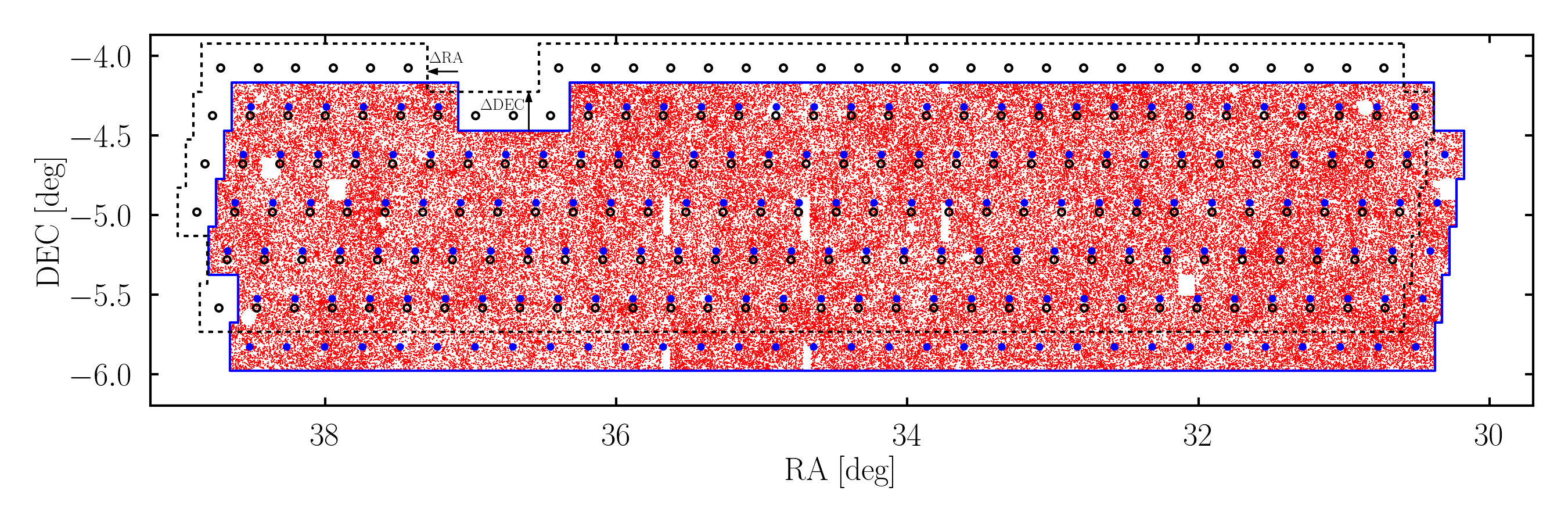}
        \includegraphics[scale=0.26]{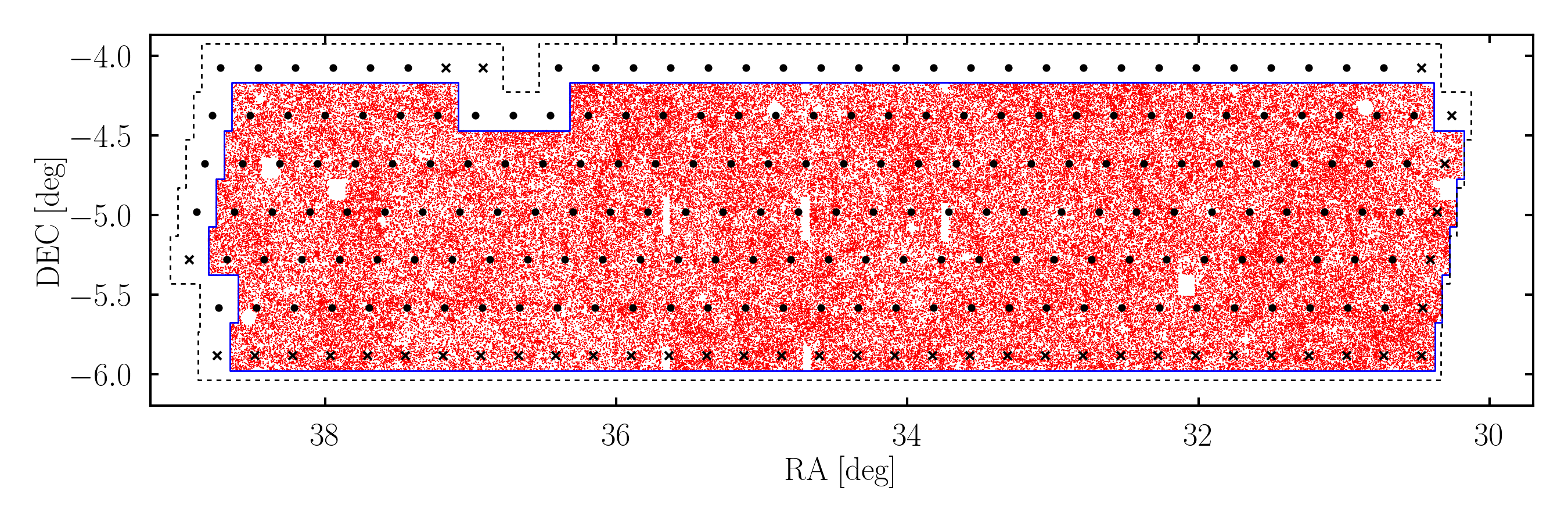}
                \caption{Top panel: Sketch showing the border effects when multiple survey realisations are generated shifting the original VIPERS spectroscopic mask over the underlying parent catalogue (red dots). The area covered by the actual VIPERS spectroscopic mask is delimited by the blue continuous line while the corresponding VIMOS pointings are displayed as blue filled dots. A random shift of $(\Delta\mathrm{RA},\Delta\mathrm{DEC})$ is then applied to obtain a new survey realisation. The area covered in the new realisation is shown as black dashed contour with black empty circles being the new positions of VIMOS pointings. We highlight the portion of the parent catalogues at low $\mathrm{RA}$ and low $\mathrm{DEC}$ that is not covered by the shifted mask. The effect is even more severe for the realisations obtained rotating the underlying parent catalogue. Bottom panel: As in the top panel but here the new survey realisation is generated shifting the `extended' spectroscopic mask (see Sect. \ref{sec:pipeline}). Black dots show the shifted position of the pointings in the original VIPERS spectroscopic mask while the black crosses represent the `artificial' pointings in the extended spectroscopic mask. The extended mask is large enough to fully cover also the parent catalogue rotated by $90$, $180,$ or $270$ degrees. In both panels a number of pointings in the shifted mask are located outside the boundaries of the parent sample. These are the pointings that only partially overlap with the parent catalogue.}\label{fig:mask_effect}
\end{figure*}

 \begin{figure}
        \centering
                \includegraphics[scale=0.20]{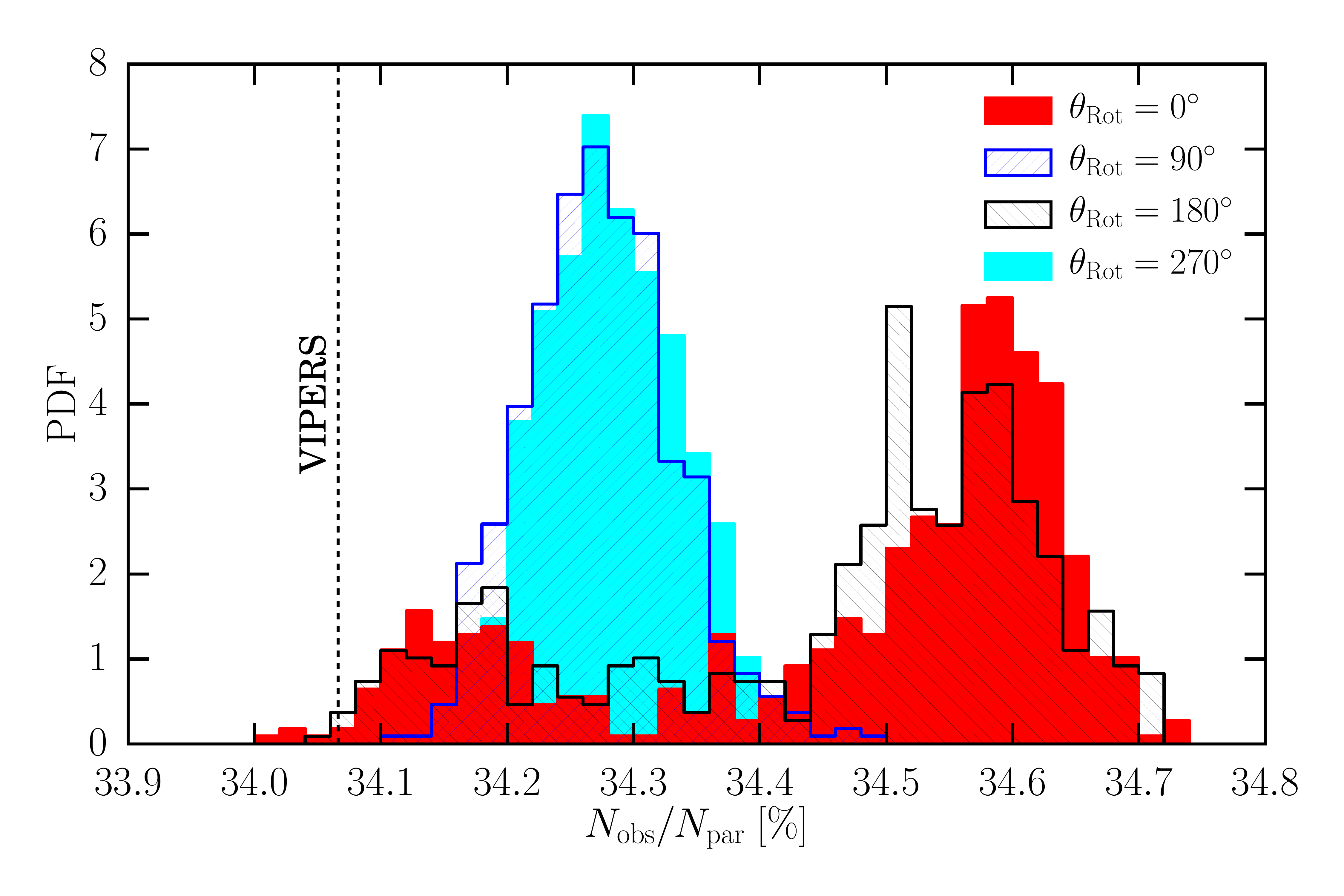}
                \caption{Normalised distributions of the observed fraction of VIPERS parent galaxies among 2170 targeting runs. Different colour coding and line styles differentiate runs with different rotation angles of the parent catalogue. The vertical dashed line shows the fractions of galaxies in the VIPERS-PDR2 galaxy catalogue.}\label{fig:targ_frac_pdf}
        \end{figure}

The weighting scheme presented in Sect. \ref{sec:pip} relies on generating multiple survey realisations to assign selection probabilities and correct the pair counts. In principle, for a slit or fiber assignment scheme that randomly selects targets in the presence of collided objects, this can be achieved by simply re-running the targeting algorithm $N_{\mathrm{runs}}$ times on the parent catalogue, with different random selection choices each time.

As described in Sect.~\ref{sec:survey}, SPOC applies a deterministic algorithm to maximise the number of slits assigned to potential targets, with no free parameters. Re-running the targeting algorithm with the same configuration of parent sample and spectroscopic mask would produce exactly the same outcome. We therefore generated multiple realisations of the spectroscopic observations from a given parent catalogue by spatially moving the spectroscopic mask in the $(\mathrm{RA,DEC})$ plane. As the VIPERS fields are equatorial, we can accurately quantify small shifts in the survey position using $\Delta$RA and $\Delta$DEC. Given the periodicity in the pattern of pointings in the VIPERS spectroscopic mask, the amount of this shift, with respect to the original VIPERS configuration, was taken as being smaller than the size of a single VIMOS pointing. We generated $N_{\mathrm{runs}}=2170$ VIPERS target realisations on each parent sample. The first of these 2170 such runs was kept fixed to the actual VIPERS-PDR2 position.

The VIPERS spectroscopic mask is defined only over the area covered by the actual VIPERS observations. A shift would therefore inevitably yield galaxies at the edges of the sample to be covered by a lower number of targeting runs with respect to those located near the centre (Fig. \ref{fig:mask_effect}). Rather than having to keep track of this, we replicated the grid of VIPERS pointings beyond the survey area such that in each run, all portions of the parent catalogue are covered by a VIMOS pointing. However, unlike the pointings in the original spectroscopic mask, we do not know the exact shapes of the quadrants belonging to the `artificial' pointings outside the survey area, so we used the shapes of the quadrants in the original VIPERS spectroscopic mask as templates and randomly assigned them to the artificial pointings. We henceforth refer to the new mask as the `extended spectroscopic mask'.

Only shifting the extended spectroscopic mask by small offsets with respect to the parent sample would require a very large number of targeting runs to accurately infer the selection probabilities and reach sub-percent level accuracy on the measurements of the two-point correlation function (see Fig. \ref{fig:un-targ_frac}). In particular, after $N_{\mathrm{runs}}=2170$ targeting runs obtained by only shifting the extended spectroscopic mask, $\sim 0.6\%$ of parent galaxies remain unobserved in any of these realisations and therefore cannot be assigned a targeting probability (Red empty circles in Fig. \ref{fig:un-targ_frac}). This fraction increases to $\sim 5.5 \%$ for galaxy-galaxy pairs (Blue filled points in Fig. \ref{fig:un-targ_frac}). This is due to the fact that under particular conditions such as in very close pairs, SPOC systematically selects the same objects in different targeting runs. This effect is quantified by the 2D angular completeness function of the sub-sample of observable pairs (i.e. the ones that are observed in any of the 2170 targeting runs) in the $(\mathrm{RA,DEC})$ plane,
        \begin{equation}\label{eq:completeness}
                C\(\mathrm{RA,DEC}\) = \frac{1+w_{\mathrm{targ}}\(\mathrm{RA,DEC}\)}{1+w_{\mathrm{par}}\(\mathrm{RA,DEC}\)} \; ,
        \end{equation}
where $w_{\mathrm{par}}$ and $w_{\mathrm{targ}}$ are the 2D angular correlation functions of the parent catalogue and its sub-sample of observable pairs, respectively. We are unable to assign selection probabilities to a significant fraction of pairs at separations $\Delta\mathrm{RA}\lesssim5''$ and $\Delta\mathrm{DEC}\lesssim130''$ due to a combination of `slit-' and `spectra-collision' as illustrated in the left panel of Fig. \ref{fig:ang-compl}.

Given the geometry of the problem, we were able to reduce the fraction of unobservable galaxies and galaxy-galaxy pairs by rotating the parent catalogue by $90^\circ$, $180^\circ$ and $270^\circ$ around an axis that passes through the sample, together with random shifts of the extended spectroscopic mask. 
Each of the 2170 survey realisations is now characterised by a rotation angle of the corresponding parent catalogue (namely $0^\circ$, $90^\circ$, $180^\circ$ and $270^\circ$) and a shift of the extended spectroscopic mask in the $(\mathrm{RA,DEC})$ plane. We stress here that we only rotate the parent sample while keeping the orientation of quadrants and dispersion direction of the galaxy spectra
fixed; that is, the larger side of the quadrants is always aligned along the declination axis. In this way we were able to assign selection probabilities to all parent galaxies and lower the fraction of unobservable galaxy-galaxy pairs to $\sim 0.06\%,$ respectively (Red dashed and blue solid lines in Fig. \ref{fig:un-targ_frac}). The price to pay is that realisations with different rotation angle of the parent catalogue are not equivalent to each other in terms of the fraction of observed galaxies. In particular, rotating the parent catalogue by $(90^\circ,270^\circ)$ provides, on average, a number of observed galaxies that is $\sim1\%$ lower than the configurations with a rotation of $(0^\circ,180^\circ)$ as shown in Fig. \ref{fig:targ_frac_pdf}. This is a consequence of the rectangular nature of the projected spectra and their alignment with the survey boundaries. This produces a different normalisation factor between these two sets of configurations that can be mitigated by angular up-weighting the pair counts. 

Given the limited number of survey realisations we used to infer selection probabilities a small fraction of pairs remain unobserved in any realisation (we refer to them as unobservable).
This introduces a systematic bias on small scales, which we do not use for RSD fitting.
As discussed in Sect. \ref{sec:pip}, given the nature of the unobservable pairs, it is appropriate to replace $DD^{(p)}(\theta)$ and $DD^{(p)}(\theta)$ in Eqs. \eqref{eq PIPauw} and Eq. \eqref{eq PIPauw DR} with $DD^{(\mathrm{targ.})}(\theta)$ and $DR^{(\mathrm{targ.})}(\theta)$, the number of observable galaxy-galaxy and galaxy-random pairs, respectively. In the following part, we use these quantities to compute the angular weights. Unless specified otherwise, we use the parent catalogue as a reference to estimate the systematic biases.

We treated the unobserved pointings and individual quadrants as a property of the photometric mask. Finally, we regarded the sky regions obscured by the photometric mask as a feature of the parent catalogues and imprinted the empty gaps accordingly. In particular, not imprinting the empty gaps due to unobserved pointings and quadrants in the parent catalogue would introduce a difference in the mean number of observed galaxies in different subsets of targeting runs. Indeed the gaps due to unobserved pointings in the uppermost row in the W1 field or in general those located far from the rotation axis would not be present in the configurations characterised by a rotation of the parent catalogue by $90^\circ$ or $270^\circ$. 

Finally, we constructed the random sample by matching the radial distribution of the VIPERS sample and imprinting the angular selection function of the parent galaxy sample, that is, applying the photometric mask. The correction scheme based on up-weighting individual galaxies according to the local densities of parent and targeted galaxies such as the TSR weighting used in \cite{delatorre13a} would have required including also the effect of the VIPERS spectroscopic mask. However, in our case this is not necessary, as this effect is already accounted for by using the PIP weighting. Including such a selection effect also in the random catalogue would have resulted in overweighting the pair counts.

\begin{figure*}
        \centering
                \includegraphics[scale=0.20]{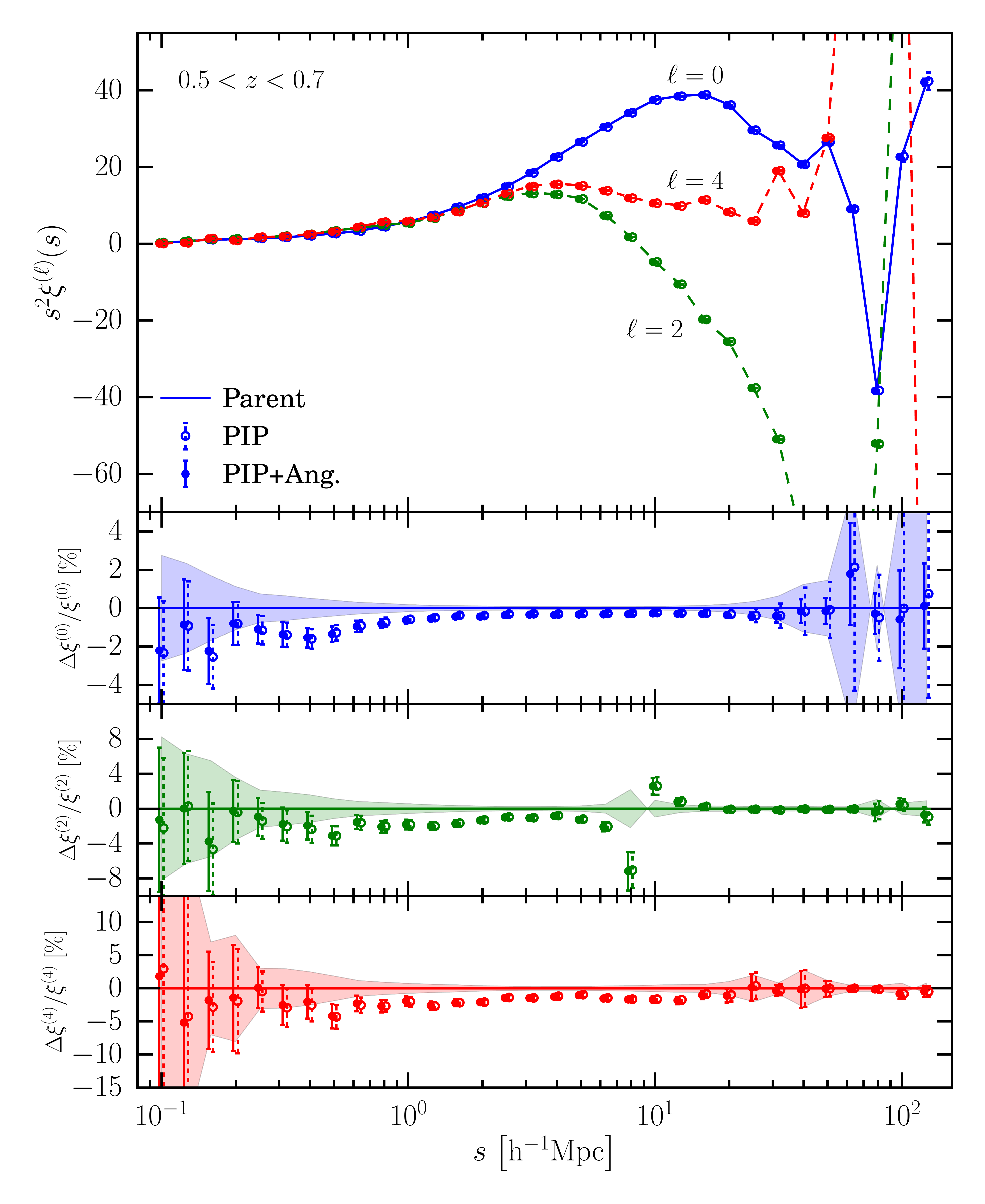}
                \includegraphics[scale=0.20]{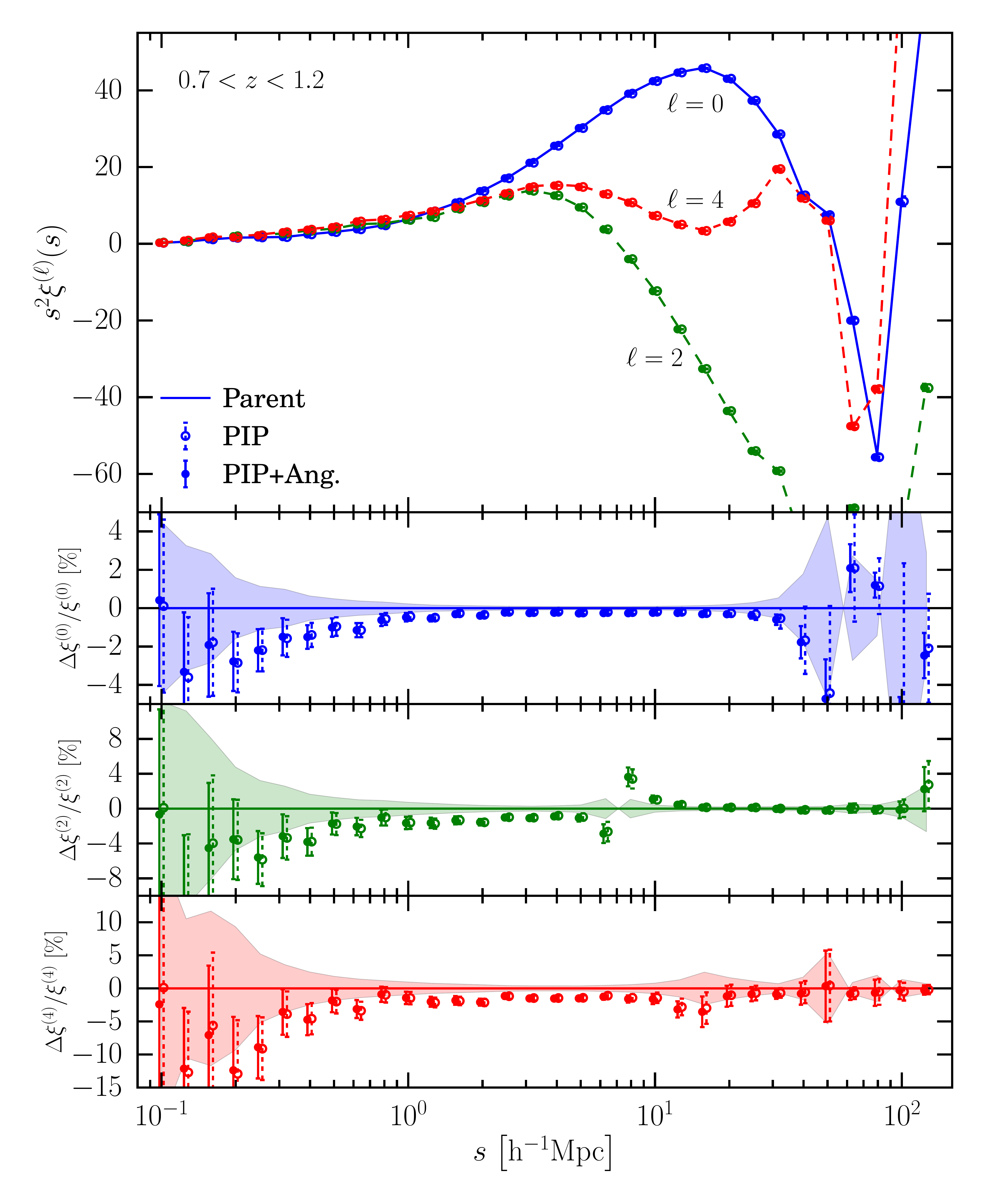}
                \caption{Top large panels: Measurements of the first three even multipoles of the two-point correlation function from one reference mock parent  sample (lines). Points with error bars show the mean and related errors among 2170 measurements obtained using the PIP weighting scheme alone (empty markers, dashed error bars) and when supplemented with an angular up-weighting (filled markers, continuous error bars) on independent survey realisations drawn from the same mock parent sample. Bottom small panels: Empty and filled markers display the fractional systematic bias of the corresponding measurements in the top large panels with respect to that from the reference mock parent sample. The horizontal continuous coloured lines and the shaded bands show the equivalent of the empty markers in the same panels but when the reference sample is limited to the galaxies and galaxy pairs that are targeted at least once in the  2170 survey realisations. Error bars in the bottom panels are obtained using the standard error propagation formula. Left and right panels show results from the lower- and higher-redshift bins, respectively. All measurements use data from W1 and W4 (mock) fields. 
        } \label{fig:pip+ang_one-mock}
        \end{figure*}

\section{Validation on mock catalogues} \label{sec:obs_syst}

\begin{figure*}
        \centering
                \includegraphics[scale=0.25]{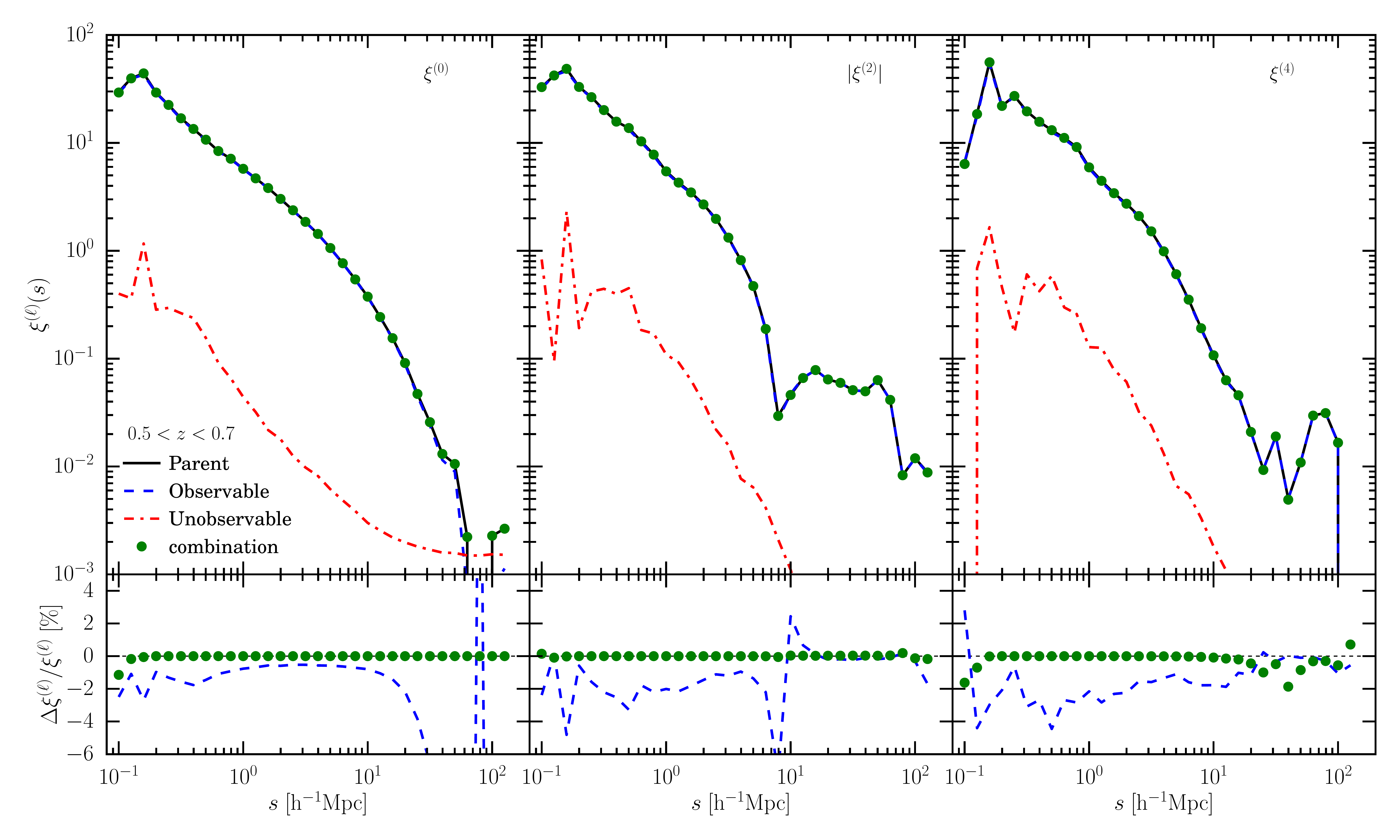}
                \caption{Top panels: Multipole moments measured from one mock parent catalogue (black continuous lines). The contribution to the overall clustering from the sub-samples of observable (blue dashed lines) and unobservable (red dash-dotted lines) pairs (defined respectively as those targeted at least once and the ones never targeted in the ensemble of 2170 targeting runs) as written in Eq. \ref{eq:LS1} are also shown. The combination of these two contributions is plotted as green filled markers. Bottom panels: Fractional offset of the contribution from observable pairs and the unobservable/observable combination with respect to the reference measurement from the parent mock. This measurement refers to the low-redshift bin $0.5<z<0.7$. The measurement in the high-redshift bin $0.7<z<1.2$ shows a very similar behaviour.} \label{fig:pip_targ+un-targ}
        \end{figure*}

\subsection{Consistency tests}

We measured the multipole moments of the two-point correlation function from each of the 2170 survey realisations, obtained by rotating the parent catalogue and shifts of the extended spectroscopic mask, using the weighted pair counts. Each of 2170 measurements was then compared to the reference estimate obtained from the mock parent catalogue to assess the mean systematic bias and related error. These measurements are shown in the top large panels of Fig. \ref{fig:pip+ang_one-mock} for the two redshift bins, while the bottom smaller panels show the corresponding fractional systematic bias with respect to the reference measurement. In particular, the PIP weighting scheme performs well over all scales with a systematic bias confined to the sub-percentage level on scales $s>1\mhmpc$ for all multipole moments in both redshift bins. These results are confirmed also when including the angular weights that however improve the statistical precision of the measurements.

The very small residual offset between the reference and the mean estimate among the corresponding 2170 realisations obtained using weighted pair counts is produced by the finite number of targeting runs that are used to sample the selection probabilities. A small fraction of galaxy-galaxy pairs is not observed in any of the targeting runs as shown in Fig. \ref{fig:un-targ_frac}. We are therefore unable to assign selection probabilities to these objects. In particular, we can split the correlation function into two summands,
        \begin{equation}
                \xi\left(\mathbf{r}\right) = \left[\frac{DD_{\mathrm{{obs}}}-2DR_{\mathrm{{obs}}}}{RR}+1\right]+\left[\frac{DD_{\mathrm{{unobs}}}-2DR_{\mathrm{{unobs}}}}{RR}\right], \label{eq:LS1}
        \end{equation}
where the first bracket represents the contribution from the subset of observable pairs while the second one results from the unobservable pairs. We measured these quantities from a mock sample using the set of corresponding bitwise weights. It is clear from Fig. \ref{fig:pip_targ+un-targ} that the unobservable pairs cluster in a very different way with respect to the galaxies in the full parent sample. They provide a non-negligible contribution to the overall clustering signal such that the expectation value of the estimator becomes different from that of the underlying parent sample. Indeed, the mean estimate of the two-point correlation function among 2170 survey realisation is unbiased if we limit the reference sample to only observable pairs.

 \begin{figure*}
        \centering
                \includegraphics[scale=0.20]{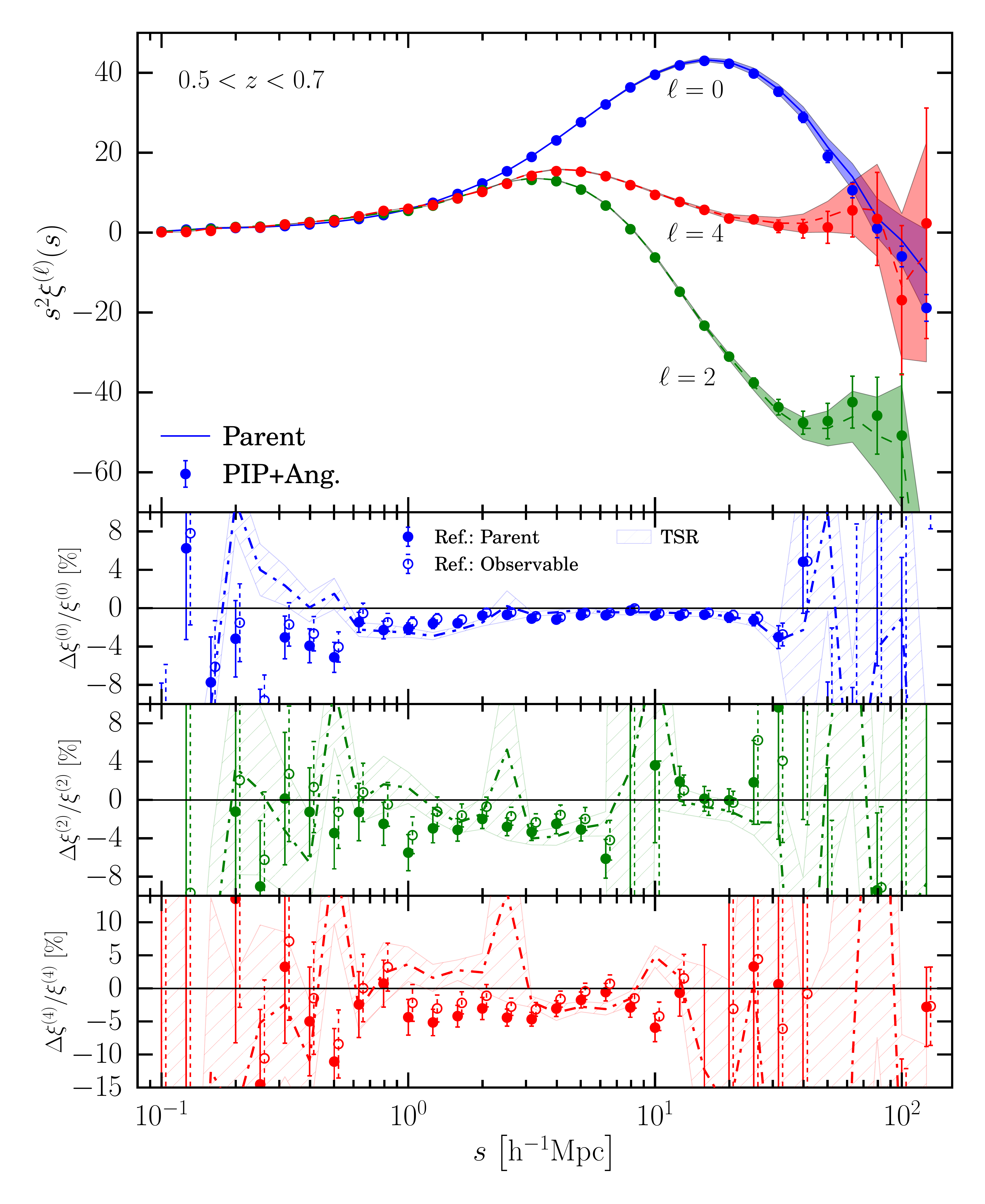}
        \includegraphics[scale=0.20]{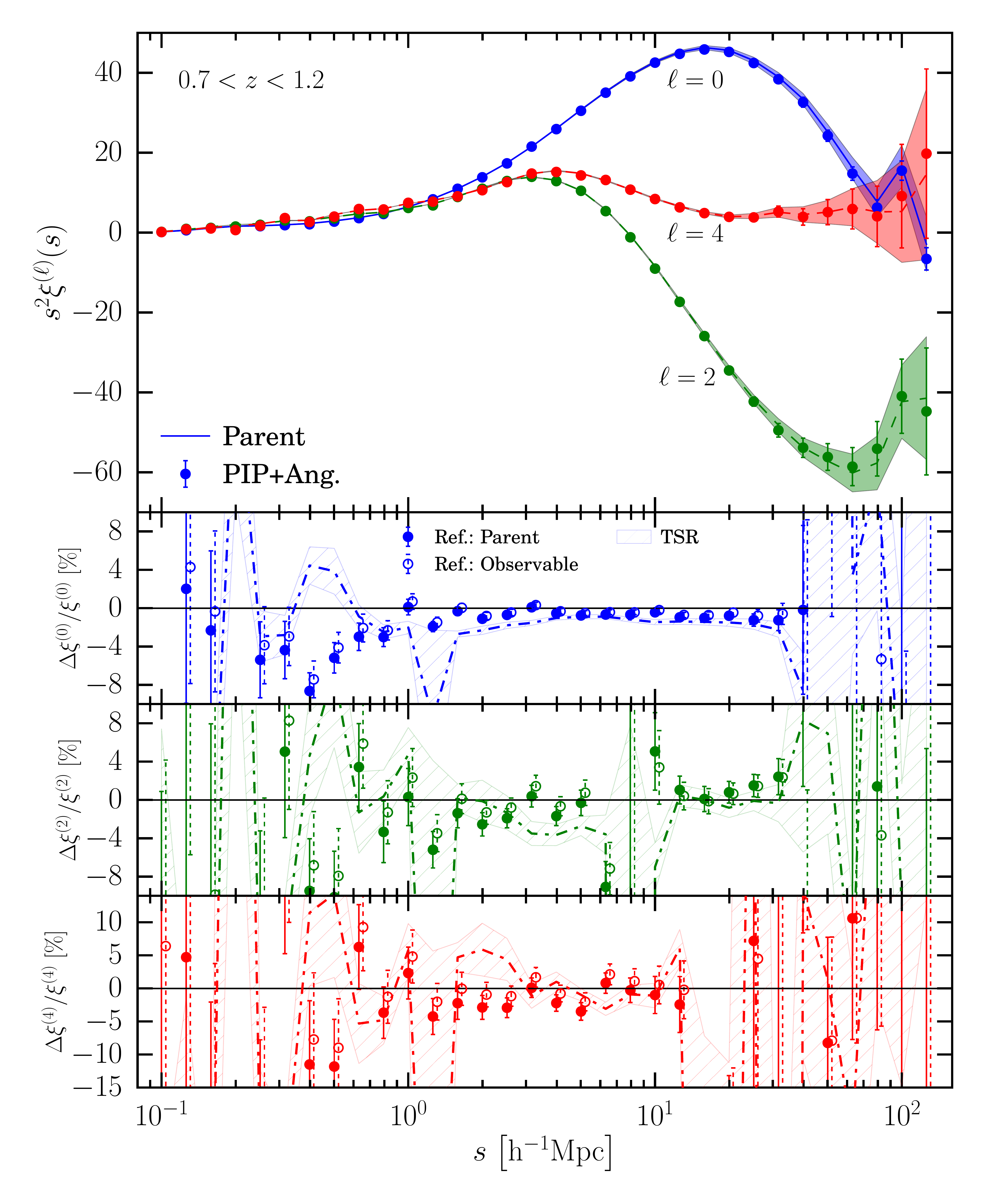}
                \caption{Top panel: Mean estimates and related errors of multipole moments of the two-point correlation function from the set of 153 mock parent samples (lines with shaded bands) and the corresponding VIPERS-like mocks obtained using the PIP and angular up-weighting method (points with error-bars). Bottom panels: Mean fractional systematic bias of measurements from VIPERS-like mocks with respect to the ones from the underlying parent samples (filled points with solid error-bars). In the bottom small panels we also display the case when the sub-sample of observable pairs is used as reference (empty markers with dashed error-bars). Measurements obtained using the TSR weighting scheme are plotted for comparison (dash-dotted lines with hatched areas). Error bars in the bottom panels quantify the scatter of the systematic offsets among 153 mocks. Left and right panels show the measurements in the low $0.5<z<0.7$ and high $0.7<z<1.2$ redshift bins, respectively.} \label{fig:pip+ang_multi-mock}
        \end{figure*}

\subsection{Observational systematic bias}

We quantified the observational systematic bias in the case where a set of 153 independent parent mocks is available and we have access to only one realisation of the spectroscopic observations for each parent sample, namely the one that matches the VIPERS-PDR2 observational configuration. We refer to this particular realisation as the VIPERS-like mock catalogue. We implemented the pipeline described in Sect. \ref{sec:pipeline} for each of the 153 mock parent samples to assign selection probabilities. We measured the multipole moments from each VIPERS-like mock using the angular up-weighting and compared to the reference measurement from the corresponding parent mock to assess the observational systematic bias. The mean and related errors on such systematic biases among 153 mocks are displayed in the bottom small panels of Fig. \ref{fig:pip+ang_multi-mock} while the corresponding mean estimates and errors among 153 parent and VIPERS-like mocks are shown in the top large panels of the same figure. The measurements from the low- and high-redshift bins are shown in the left and right panels, respectively.

The new weighting scheme provides clustering measurements accurate at the sub-percentage level down to very small scales ($\sim 0.6\mhmpc$) in both redshift ranges. The systematic bias increases at scales of $\gtrsim40\mhmpc$ remaining within $2\sigma$ of the reference estimates. The residual systematic offset on scales of interest ($\lesssim50\mhmpc$) results from a combination of two effects: a) in each mock sample a small fraction of galaxy pairs remain unobserved in the ensemble of 2170 survey realisations (see Fig. \ref{fig:un-targ_frac} and Fig. \ref{fig:pip_targ+un-targ}); b) the VIPERS-like configuration is not a random realisation but rather a particular case among the  2170 survey realisations used to infer selection probabilities, namely the one characterised by a rotation angle of the parent sample of $\theta=0^\circ$ and no shifts (see Fig. \ref{fig:targ_frac_pdf}).
Figure \ref{fig:pip+ang_multi-mock} also shows results obtained up-weighting each galaxy by the corresponding target sampling rate (TSR). This technique, used in  previous analyses of VIPERS-PDR2 data, performs similarly to the new method tested in this work. It is important to recall here that the TSR weighting scheme was calibrated to minimise the systematic bias on clustering estimates in mock catalogues. As such it does not assure a similar performance on real data due to possible differences between the clustering of real and simulated galaxies.

\section{RSD Fitting}           \label{sec:rsd_mocks}    

\subsection{Theoretical modelling}

We modelled the anisotropic clustering in the monopole $\xi^{(0)}$ and quadrupole $\xi^{(2)}$ two-point correlation functions as described in \citet{pezzotta16}. We used 
the TNS model \citep{taruya10} that reads in the case of biased tracers,
\begin{equation}
        \begin{split}
        P^s\(k,\mu_k\) =        &D\(k\mu_k\sigma_{v}\)[b^2P_{\delta\delta}\(k\)+2\mu_k^2fbP_{\delta\theta}+\mu_k^4f^2P_{\theta\theta}\(k\)+\\
                                        & A\(k,\mu_k,f,b\) + B\(k,\mu_k,f,b\)]\ ,\label{eq:taruya}
        \end{split}     
\end{equation}
with $f$ and $b$ being the growth rate and linear galaxy bias, respectively. In Eq. \eqref{eq:taruya}, $P_{\delta\delta}$ is the non-linear matter power spectrum and $P_{\delta\theta}$ and $P_{\theta\theta}$ are density-velocity divergence and velocity divergence-velocity divergence power spectra, respectively. The correction factors $A\(k,\mu_k,f,b\)$ and $B\(k,\mu_k,f,b\)$ are derived using perturbation theory and provided in \cite{taruya10} and \citet{delatorre12}, and account for the mode coupling between density and velocity fields. The phenomenological damping factor $D(k\mu_k\sigma_{v})$ mimics the effect of the small-scale pairwise velocity dispersion by suppressing the clustering power predicted by the `Kaiser factor' and depends on the nuisance parameter $\sigma_{v}$. We used a Lorentzian functional form for $D(k\mu_k\sigma_v)$ as it is found to better describe the observations with respect to the theoretically predicted Gaussian damping factor \citep[e.g.][]{pezzotta16}. The model in Eq. \ref{eq:taruya} is also supplemented with a second Gaussian damping factor with fixed dispersion $\sigma_z$ to account for the effect of VIPERS redshift errors on clustering measurements.

The model in Eq. \ref{eq:taruya} depends on four fitting parameters $(f,b,\sigma_8,\sigma_{v})$. However, we provide measurements of the derived parameters $f\sigma_8$ and $b\sigma_8$ as $\sigma_8$, the normalisation of the linear matter power spectrum $P_{\delta\delta}^{\mathrm{lin}}$, is degenerate with the growth rate parameter $f$ and the linear bias factor $b$. 

The linear matter power spectrum $P_{\delta\delta}^{\rm lin}$ is obtained using the Code for Anisotropies in the Microwave Background \citep[][CAMB]{lewis00} that is combined with HALOFIT \citep{smith03,takahashi12} to predict the non-linear matter power spectrum $P_{\delta\delta}$. The density-velocity divergence $P_{\delta\theta}$ and velocity divergence-velocity divergence $P_{\theta\theta}$ power spectra cannot be measured from data directly. They can be predicted by either using perturbation theory or by means of empirical fitting functions calibrated on numerical simulations \citep[e.g.][]{jennings11}. Perturbation theory however breaks down at scales accessible in VIPERS. We therefore used the improved fitting functions described in \citet{bel17},
\begin{subequations}
                \label{eq:fitting}
                \begin{align}
                        \pdt            &=\[P_{\delta\delta}^{\rm{lin}}\(k\)P_{\delta\delta}\(k\)\exp\(-\frac{k}{k_{\delta\theta}^{\rm{cut}}}\) \]^{1/2}\; ,\\
                        \ptt            &=\[P_{\delta\delta}^{\rm{lin}}\(k\)\exp\(-\frac{k}{k_{\theta\theta}^{\rm{cut}}}\)\]\; .
                \end{align}
        \end{subequations}
In Eq. \eqref{eq:fitting}, $k_{\delta\theta}^{\rm{cut}}$ and $k_{\theta\theta}^{\rm{cut}}$ are defined as
        \begin{subequations}
                \label{eq:kcut}
        \begin{align}
                        k_{\delta\theta}^{\rm{cut}}           &=\frac{1}{2.972}\sigma_8^{-2.034}\; ,\\
                        k_{\theta\theta}^{\rm{cut}}           &=\frac{1}{1.906}\sigma_8^{-2.163}\; ,
                \end{align}
        \end{subequations}
with $\sigma_8$ being the amplitude of the linear matter power spectrum. We note that in our model, $\sigma_8$ controls the level of non-linearity (within HALOFIT) in the matter non-linear density-velocity divergence and velocity divergence-velocity divergence power spectra that enter the RSD model of Eq. \ref{eq:taruya}.

\subsection{Fitting method and data covariance matrix}          \label{sec:cov} 
 \begin{figure}
        \centering
                \includegraphics[scale=0.12]{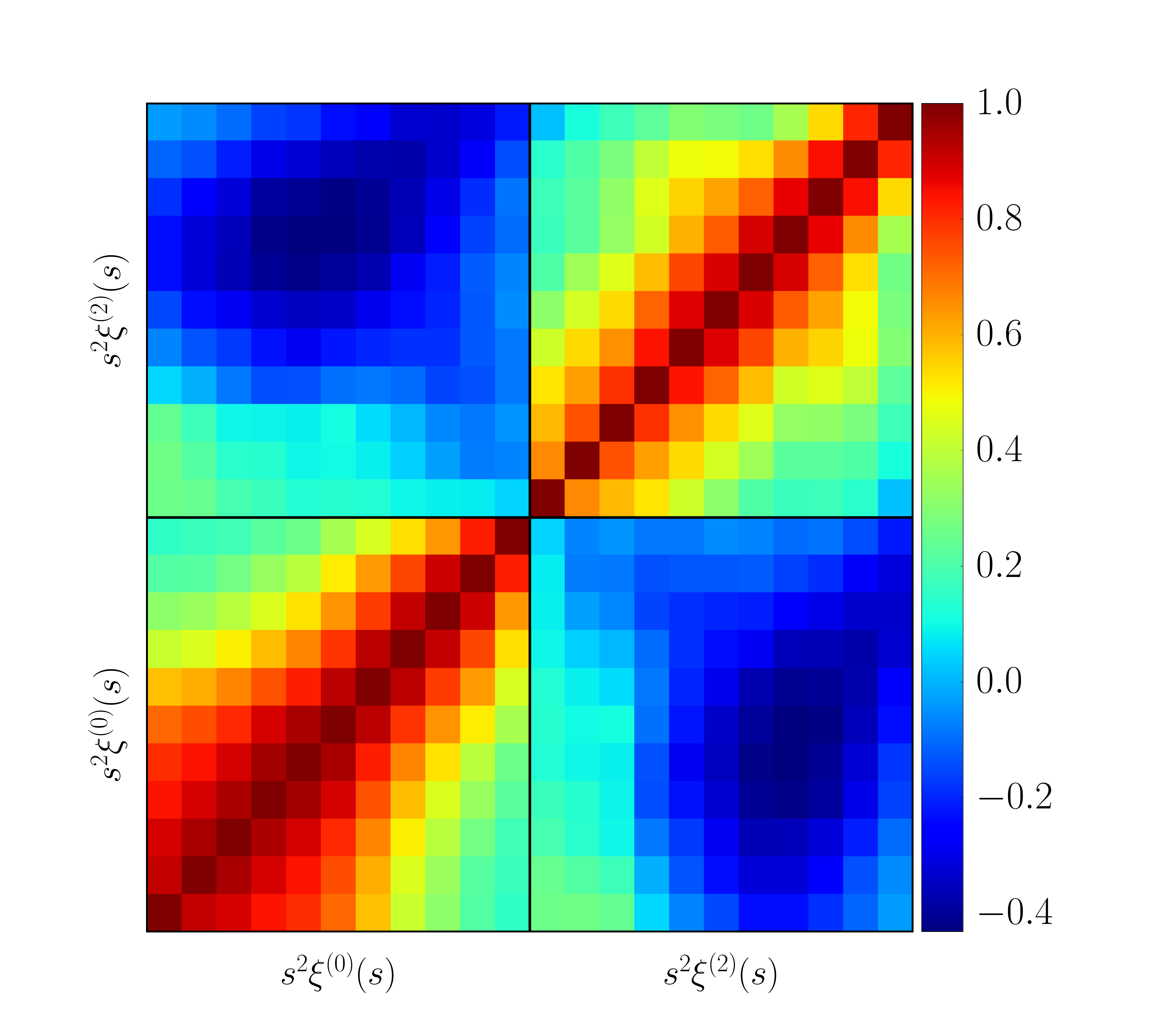}
        \includegraphics[scale=0.12]{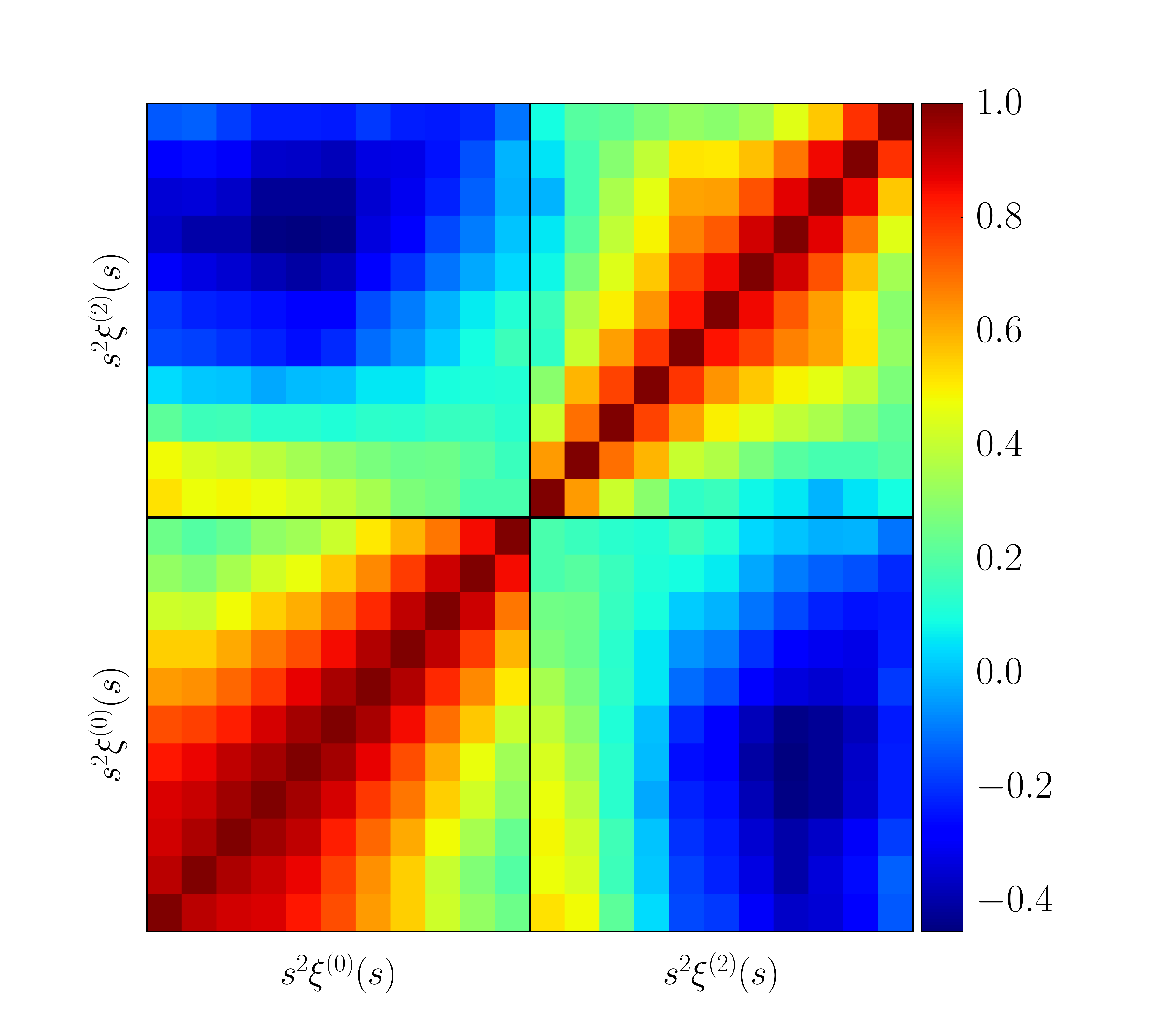}
                \caption{Data correlation matrices $R_{ij}=C_{ij}/\sqrt[]{C_{ii}C_{jj}}$ estimated using the set of 153 VIPERS-like mock catalogues in the low- $0.5<z<0.7$ (top panel) and high-redshift bins $0.7<z<1.2$ (bottom panel).} \label{fig:corr_mat}
        \end{figure}

 \begin{figure}
        \centering
                \includegraphics[scale=0.62]{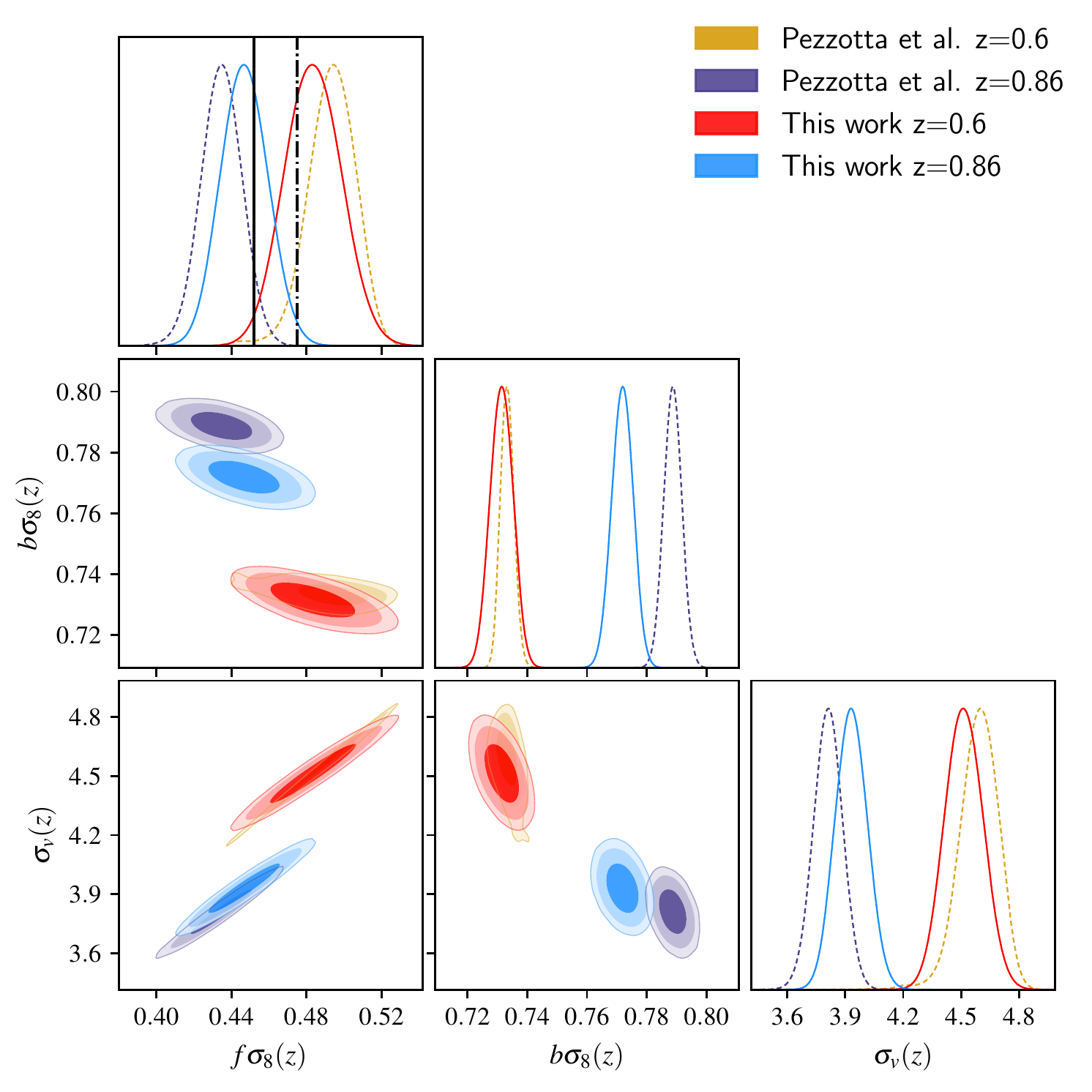}
                \caption{One- and two-dimensional marginalised posterior likelihoods of the derived parameter $f\sigma_8$, $b\sigma_8$ and the nuisance parameter $\sigma_{v}$ resulting from the analysis of the mean clustering estimates obtained from 153 VIPERS-like mock catalogues using the method in Sect. \ref{sec:pip}. Fits are performed with TNS model between a minimum fitting scale of $s_{\mathrm{min}}=5\mhmpc$ up to a maximum scale of $s_{\mathrm{max}}=50\mhmpc$. For comparison we have also over-plotted results obtained in \citet{pezzotta16} using the same set of mock samples and fitting method. Vertical dash-dotted and solid lines correspond to the expected values of $f\sigma_8$ at $z=0.6$ and $z=0.86,$ respectively.} \label{fig:plike_mocks}
        \end{figure}

The measured monopole and quadrupole are simultaneously fitted with the TNS model to estimate the fitting parameters using the Monte-Carlo Markov Chain (MCMC) technique. The MCMC algorithm explores the posterior distribution in the parameter space constrained by the data likelihood and parameter priors. The data likelihood is,
\begin{equation}
        -2\ln\mathcal{L} = \chi^2\(\theta_p\) = \sum_{i,j}\Delta_i\(\theta_p\)\ C_{ij}^{-1}\ \Delta_j\(\theta_p\)       \; ,            \label{eq:chi}
\end{equation}
where $\theta_p$ denotes the set of fitting parameters, $\Delta_i$ is the discrepancy between the data and model prediction in bin $i$ and $C_{ij}^{-1}$ is the precision matrix, that is, the inverse of the data covariance matrix $C_{ij}$. We fit the monopole $s^2\xi^{(0)}$ and quadrupole $s^2\xi^{(2)}$ of the two-point correlation functions simultaneously and accounted for their cross-covariance in the data covariance matrix.

The covariance matrices $C_{ij}$ were estimated using the set of 153 VIPERS-like mocks. Noise in the covariance matrix is amplified when inferring the precision matrix using $C_{ij}$ and leads to a biased estimate of the precision matrix. We corrected for this bias by means of the corrective factor provided in \cite{percival14}. The correlation matrices, that is, $R_{ij}=C_{ij}/(C_{ii}C_{jj})^{1/2}$, for the two redshift bins and restricted to the range of fitting scales used here are shown in Fig. \ref{fig:corr_mat}.

The robustness of the data analysis method has already been tested in \citep{pezzotta16}. We therefore focus on repeating the analysis using only the range of fitting scales adopted in \citet{pezzotta16} to obtain the reference estimates of the $f\sigma_8$ parameter, that is, minimum and maximum scales fixed at $s_{\mathrm{min}}=5\mhmpc$ and $s_{\mathrm{max}}=50\mhmpc$, respectively. In particular, we fit the mean estimates of $s^2\xi^{(\ell)}$ for the monopole $\ell=0$ and quadrupole $\ell=2$ from the mock catalogues with the TNS model and obtained a systematic offset, with respect to the fiducial values, of
\begin{subequations}
        \begin{align*}
                \Delta \left(f\sigma_8\right)(z=0.60)&=0.009\pm0.015\\
        \Delta \left(f\sigma_8\right)(z=0.86)&=-0.006\pm0.012 \; .
    \end{align*}
\end{subequations}
These estimates are un-biased compared to the expected values of $f\sigma_8(z)$ in the mock fiducial cosmology. Moreover our measurements are also compatible with estimates obtained in \citet{pezzotta16}, $\Delta\left(f\sigma_8\right)(z=0.60)=0.019\pm0.012$ and $\Delta\left(f\sigma_8\right)(z=0.86)=-0.018\pm0.011$, within $1\sigma$. The marginalised one- and two-dimensional posterior likelihoods are shown in Fig. \ref{fig:plike_mocks}. For comparison we also show, in the same figure, the results obtained by \citet{pezzotta16} using the same set of mocks.

\section{Growth rate measurements}      \label{sec:rsd_data}   

 \begin{figure}
        \centering
                \includegraphics[scale=0.62]{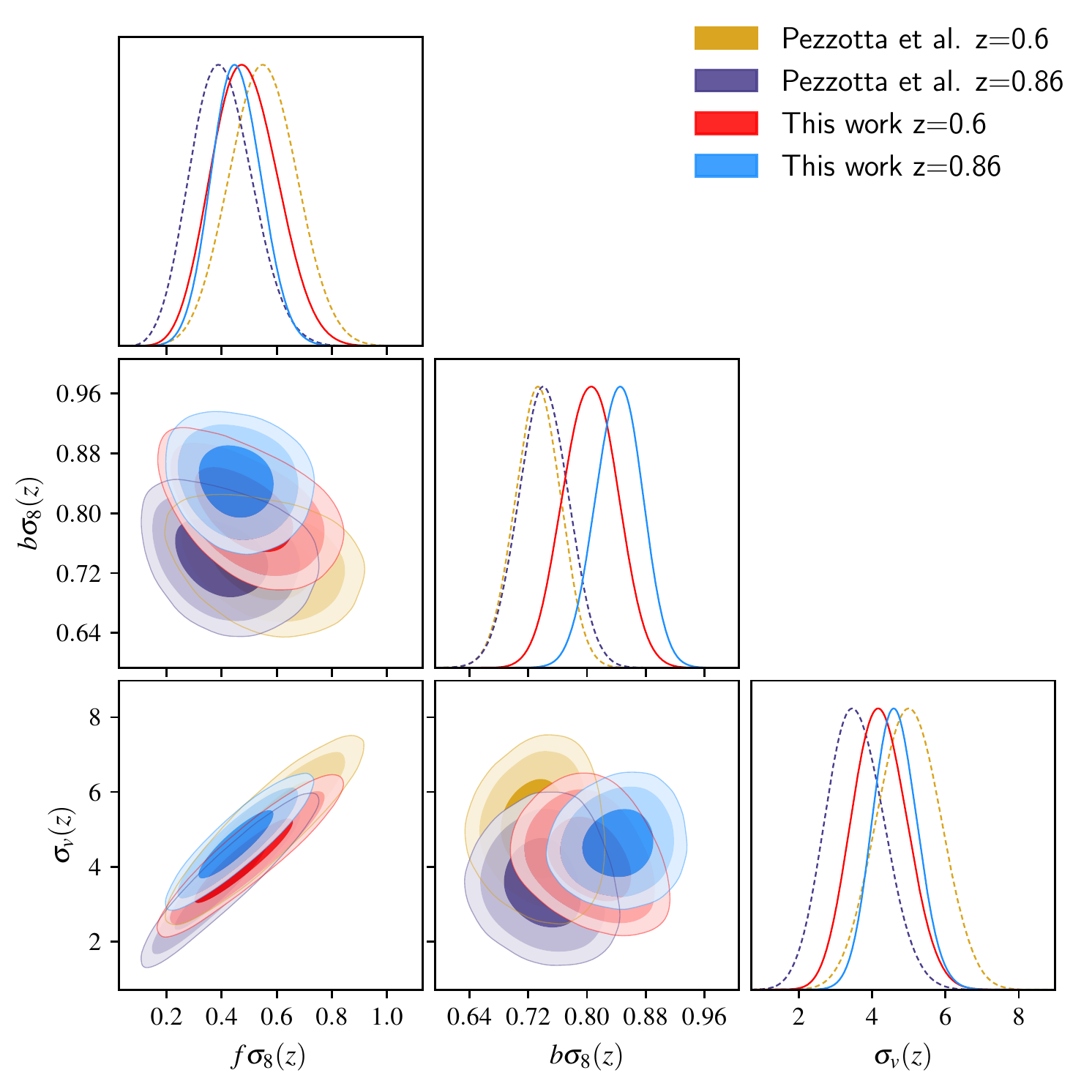}
                \caption{As in Fig. \ref{fig:plike_mocks} but now fitting the monopole $\xi^{(0)}$ and quadrupole $\xi^{(2)}$ measured from the VIPERS-PDR2 spectroscopic sample. Again, results of RSD fitting in \citet{pezzotta16} are also plotted.} \label{fig:plike_pdr2}
        \end{figure}

 \begin{figure}
        \centering
                \includegraphics[scale=0.20]{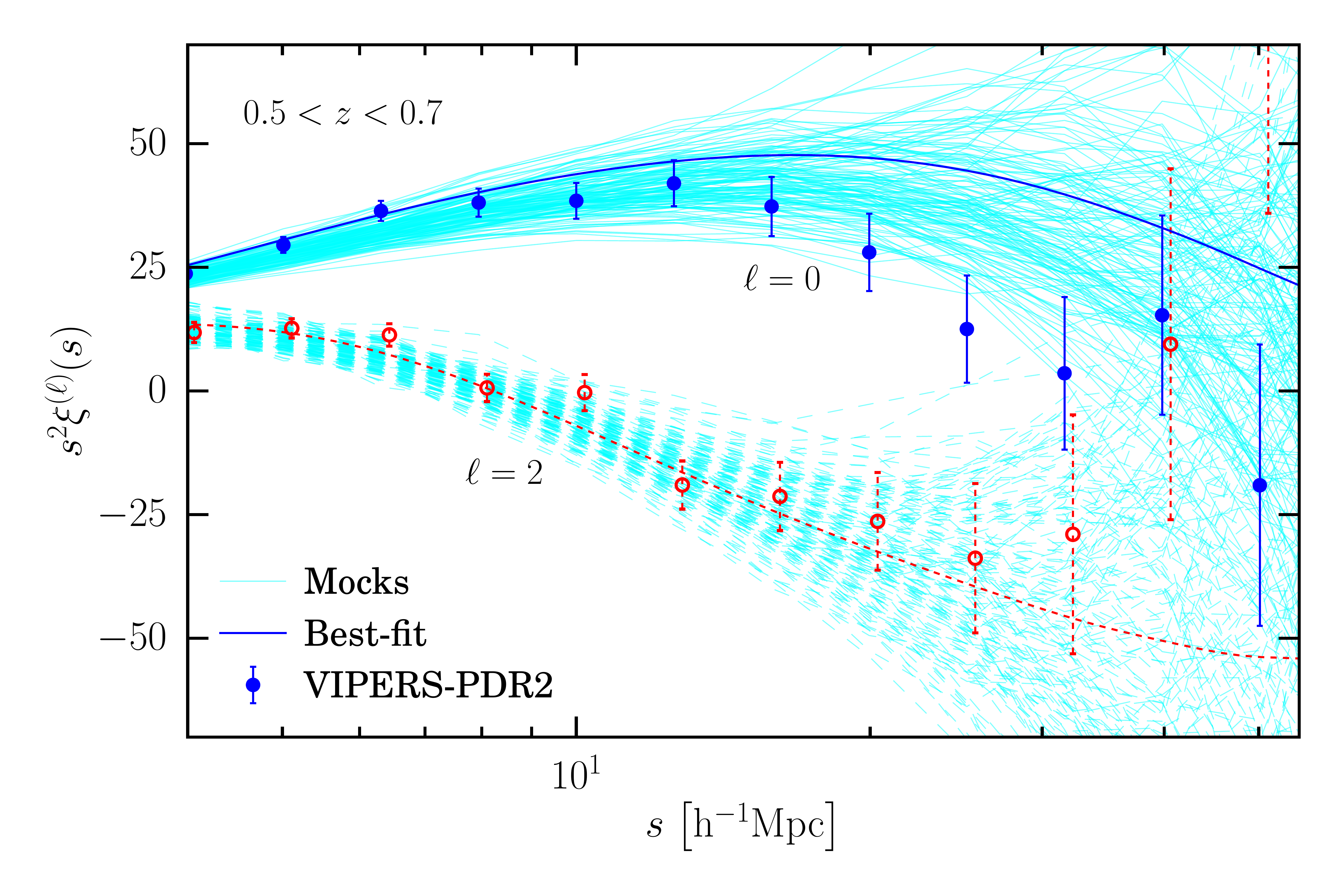}
	        \includegraphics[scale=0.20]{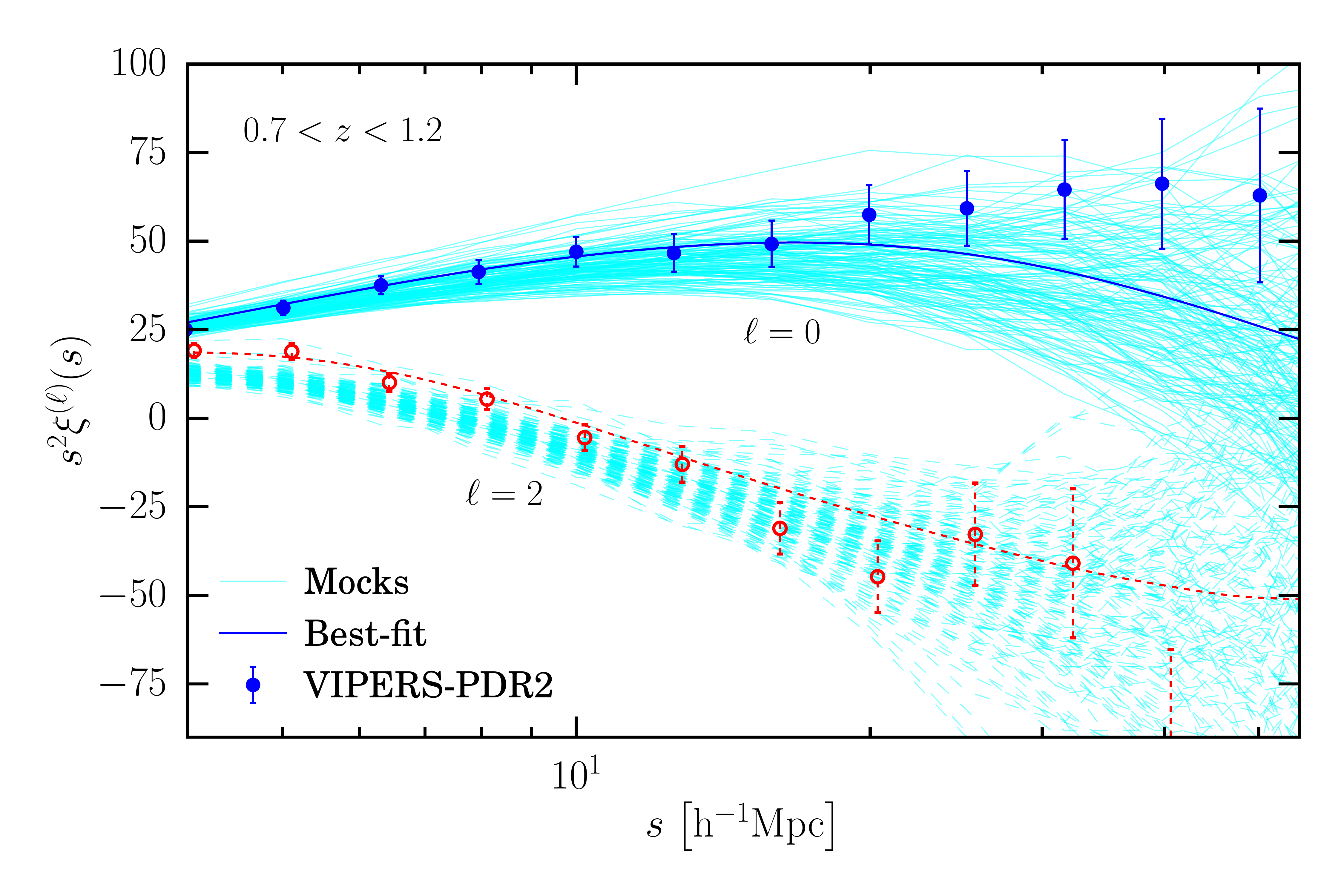}
                \caption{Monopole $s^2\xi^{(0)}$ and quadrupole $s^2\xi^{(2)}$ moments of the two-point correlation function measured from VIPERS-PDR2 galaxy sample using the weighted pair counts as described in Sect. \ref{sec:pip} (points with error-bars). Diagonal errors are estimated using the set of 153 VIPERS-like mocks. Cyan lines show the measurements from individual VIPERS-like mocks. The best-fit models corresponding to the results in Fig. \ref{fig:plike_pdr2} are also displayed as solid blue and dashed red lines. Top and bottom panels show results from the low- and high-redshift bins, respectively.} \label{fig:best-fit}
        \end{figure}

To correct the measurements of the two-point correlation function from the VIPERS-PDR2 galaxy catalogue we followed the same procedure adopted on mock catalogues, including calculating the PIP weights using both rotations to the parent catalogue and shifts of the extended spectroscopic mask. As VIPERS parent catalogue we used the photometric catalogue from the CFHTLS W1 and W4 fields, from which VIPERS targets were drawn, restricted to the area covered by the VIPERS observations. However, unlike mock samples the VIPERS parent catalogue contains $N_c=449$ compulsory targets that do not enter the maximisation of the number of slits. Although a negligible fraction, we accounted for these objects when generating multiple survey realisations unless they fall inside the empty gaps between VIMOS quadrants. As anticipated in Sect. \ref{sec:survey}, we used only galaxies with quality flags 2 to 9 (inclusive) corresponding to a sample with a redshift confirmation rate of 96.1\%. The effect of redshift failures is not accounted for when computing the PIP weights. We therefore corrected for the effect of redshift failures by up-weighting each galaxy in the VIPERS-PDR2 catalogue by the corresponding SSR as described in Sect. \ref{sec:ssr}.

We fit the monopole $s^2\xi^{(0)}$ and quadrupole $s^2\xi^{(2)}$ of the two-point correlation function between $s_{\mathrm{min}}=5\mhmpc$ and $s_{\mathrm{max}}=50\mhmpc$ with the TNS model in Eq. \ref{eq:taruya} supplemented with a second Gaussian damping factor with width fixed to the VIPERS spectroscopic redshift error $\sigma_z/(1+z)=0.00054$. The measured values for the derived parameter $f\sigma_8$ are,
\begin{subequations}
        \begin{align*}
                f\sigma_8(z=0.60)&=0.49\pm0.12\\
        f\sigma_8(z=0.86)&=0.46\pm0.09 \; .
    \end{align*}
\end{subequations}
These values are compatible within 1-$\sigma$ with estimates from \citet{pezzotta16}, namely $f\sigma_8(z=0.60)=0.55\pm0.12$ and $f\sigma_8(z=0.86)=0.40\pm0.11$, who used the same datasets and theoretical prescriptions for RSD modelling. Furthermore our measurements are also consistent within the error bars with the ones obtained with alternative methods such as a combination of RSD and galaxy-galaxy lensing in \citet{delatorre16} or the one using a sample of luminous blue galaxies in VIPERS as done in \citet{mohammad17}. The best-fit models corresponding to the results in Fig. \ref{fig:plike_pdr2} are displayed in Fig. \ref{fig:best-fit} along with the measurements of the monopole $s^2\xi^{(0)}$ and quadrupole $s^2\xi^{(2)}$ moments of the two-point correlation function using the VIPERS-PDR2 galaxy sample (points with error-bars) and VIPERS-like mocks (cyan lines).

\section{Summary and conclusions}  \label{sec:conc}

We corrected the clustering estimates from the VIPERS-PDR2 galaxy sample using the PIP method described in \citet{bianchi17}. This technique was supplemented with the angular up-weighting scheme proposed in \citet{percival17} to improve the statistical precision of the measurements. The PIP method relies on up-weighting the pair-counts based on the corresponding selection probabilities. These probabilities were inferred empirically by generating multiple survey realisations from a parent catalogue and counting the number of times a given pair is observed. To compare the performance of this new technique with the results obtained in \citet{pezzotta16} we split the redshift range probed by VIPERS into two bins spanning $0.5<z<0.7$ and $0.7<z<1.2$. The following considerations equally apply to both redshift bins.

Given the features of the VIPERS targeting algorithm and the limited extension of the VIPERS parent mocks, we generated multiple (2170) VIPERS realisations from each parent sample by spatially moving the spectroscopic mask. To assign selection probabilities to galaxy pairs with a reasonable amount of computation time we also rotated the parent catalogue in each targeting run. The price to pay is that survey realisations with different rotations of the parent sample are not fully equivalent to each other producing a `normalisation problem' for the weighted pair counts. We mitigated for this problem supplementing the PIP technique with the angular up-weighting method. A negligible mean systematic bias was found comparing clustering measurements from each of 2170 survey realisations with the reference measurement from the parent catalogue. Nevertheless, we have shown that this bias is produced by the very small fraction of galaxy pairs unobserved in $N_{\mathrm{runs}}=2170$ survey realisations. Indeed these pairs are not randomly distributed but rather exhibit a small-scale clustering.

To assess the observational systematic bias on clustering measurements, we selected, for each parent mock, only the survey realisation obtained with actual VIPERS observational setup, that is no rotation of the parent sample and no shift in the spectroscopic mask. We found a mean fractional systematic bias among 153 mock samples to be below the percentage level. We argue that such a small offset results from a combination of two effects: a) we are unable to assign selection probabilities to a small fraction of pairs that cannot be observed using only 2170 survey realisations, referred to as unobservable pairs; and b) the VIPERS-like mock is a particular configuration among the 2170 realisations used to infer the selection probabilities. Our tests using mocks catalogues have shown the new method to be a valid and robust way to correct for missing targets in VIMOS observations.

We tested the impact of these corrections on estimates of the growth rate of structure times the amplitude of dark matter density fluctuations $f\sigma_8$. In particular we fitted the mean estimates of the corrected monopole and quadrupole among 153 VIPERS-like mocks with the TNS model on scales $5\mhmpc<s<50\mhmpc$. The analysis provided un-biased estimates of the fitting parameter $f\sigma_8$ that are fully consistent with those obtained in \citet{pezzotta16} using the same configuration of fitting scales and theoretical model. The measurements made using the new technique are slightly closer to the expected values, but the difference is within the expected errors. This provides further confirmation of the robustness of previous RSD analyses in VIPERS. However we stress here the fact that while the correction scheme adopted in previous VIPERS works \citep[e.g.][]{delatorre16, pezzotta16, mohammad17} relied on a fine-tuned parametric approach calibrated on mock catalogues to minimise the observational systematic bias, the new technique proposed in \citet{bianchi17} and \citet{percival17} is exact and is self-contained, using only the data itself.

Finally, we applied this method to correct the measurements of the two-point correlation function using the VIPERS-PDR2 galaxy catalogue. When dealing with data we have accounted for the effect of redshift failures by means of the so called `spectroscopic success rate' (SSR). We also took into account the presence of a small fraction of compulsory targets in the parent sample. Both these features were not reproduced in the mock samples. The measured monopole and quadrupole moments of the two-point correlation functions were fitted with the TNS model to estimate the $f\sigma_8$ parameter at the effective redshifts of the two redshift bins. Our measurements are in agreement within 1-$\sigma$ with previous measurements by \citet{pezzotta16}, \citet{delatorre16} and \citet{mohammad17} at the same redshifts.

In future work, we will improve upon this analysis using the method of \citet{percival17} to include angular clustering measurements from the full CFHTLS sample. By using a combination of the angular and 3D clustering measurements, we hope to observe baryon acoustic oscillations, as well as to improve on the current RSD measurements.

\begin{acknowledgements}

We thank Michael Wilson for useful comments on this work. We acknowledge 
the crucial contribution of the ESO staff for the management of service 
observations through which the VIPERS survey was built. In particular, we 
are deeply grateful to M. Hilker for his constant help and support of this
programme. Italian participation to VIPERS has been funded by INAF
through PRIN 2008, 2010, 2014 and 2015 programs. 
LG, FGM, BRG and JB acknowledge support from the European Research Council 
through grant n.~291521. DB and WJP acknowledge support from the European 
Research Council through grant n.~614030. OLF acknowledges support from the 
European Research Council through grant n.~268107.
SDLT acknowledges the support of the OCEVU Labex
(ANR-11-LABX-0060) and the A*MIDEX project (ANR-11-IDEX-0001-02)
funded by the "Investissements d'Avenir" French government programme
managed by the ANR.
RT acknowledges financial support from the European Research Council through grant n.~202686. 
AP, KM, and JK have been supported by the National Science Centre (grants
UMO-2012/07/B/ST9/04425 and UMO-2013/09/D/ST9/04030). EB, FM and LM
acknowledge the support from grants ASI-INAF I/023/12/0 and PRIN MIUR
2010-2011. 
TM and SA acknowledge financial support from the ANR Spin(e) through the French grant
ANR-13-BS05-0005.

The Big MultiDark Database used in this paper and the web application providing online access to it were constructed as part of the activities of the German Astrophysical Virtual Observatory as result of a collaboration between the Leibniz-Institute for Astrophysics Potsdam (AIP) and the Spanish MultiDark Consolider Project CSD2009-00064. The Bolshoi and MultiDark simulations were run on the NASA's Pleiades supercomputer at the NASA Ames Research Center.

\end{acknowledgements}

\bibliographystyle{aa}
\bibliography{biblio_gg,biblio_VIPERS_v4}

\end{document}